\documentclass[11pt]{article}

\usepackage{ifpdf}
\ifpdf
   \usepackage[pdftex]{graphicx}
\else
   \usepackage{graphicx}
\fi
\usepackage{epstopdf}

\usepackage{amssymb}
\usepackage{latexsym}
\RequirePackage{amsmath}
\usepackage{algorithm}
\usepackage{algpseudocode}
\usepackage{color}
\usepackage{array}
\usepackage[square,numbers,sort&compress]{natbib}

\newcommand{\Valhalla}{\textsc{Valhalla}\xspace}
\newcommand{\new}[1]{{#1}}
\newcommand{\old}[1]{{}}

\input{sections/defs}

\begin{document}

\title{Block-Structured Adaptive Mesh Refinement Algorithms for Vlasov Simulation}

\author{J. A. F. Hittinger and J. W. Banks}


\maketitle 

\begin{abstract}
Direct discretization of continuum kinetic equations, like the Vlasov
equation, are under-utilized because the distribution function generally
exists in a high-dimensional ($>$3D) space and computational cost
increases geometrically with dimension.  We propose to use high-order
finite-volume techniques with block-structured adaptive mesh refinement
(AMR) to reduce the computational cost.  The primary complication comes
from a solution state comprised of variables of 
different dimensions.  We develop the algorithms required to
extend standard single-dimension block structured AMR to the
multi-dimension case.  Specifically, algorithms for reduction and
injection operations that transfer data between mesh hierarchies of
different dimensions are explained in detail.  In addition,
modifications to the basic AMR algorithm that enable the use of
high-order spatial and temporal discretizations are discussed.
Preliminary results for a standard 1D+1V Vlasov-Poisson test problem are
presented.  Results indicate that there is potential for significant
savings for some classes of Vlasov problems.
\end{abstract}


\section{Introduction} 
\label{sec:introduction}
The Vlasov-Maxwell system of equations is a fundamental kinetic model
that describes weakly-coupled plasma dynamics. The Vlasov equation is a
partial differential equation in phase space, 
$(\mbx,\mbv)\in\field{R}^{N}\!\times\!\field{R}^{M}$
for $N,M\in[1,2,3]$ such that $M\ge N$, that 
describes the evolution in time, $t\in\field{R}_+$, of a particle 
distribution function, $f(\mbx,\mbv,t)\in\field{R}_+$, in the presence 
of electromagnetic fields, $\mbE(\mbx,t)\in\field{R}^{N}$ and 
$\mbB(\mbx,t)\in\field{R}^{N}$.  For a particle species $\alpha$, the 
Vlasov equation is 
\begin{equation}
  \pderiv{f_{\alpha}}{t} + \mbv\cdot\nabla_{\mbx}f_{\alpha} 
  + \frac{q_{\alpha}}{m_{\alpha}}\left(\mbE+\mbv\times\mbB\right)
  \cdot\nabla_{\mbv}f_{\alpha} = 0,
  \label{eq:vlasov_nc}
\end{equation}
where the particle charge and mass are $q_{\alpha}$ and $m_{\alpha}$,
respectively.  Both imposed and self-generated electric and magnetic
fields, $\mbE$ and $\mbB$, respectively, are responsible for the Lorentz 
force in~\eqref{eq:vlasov_nc} and are the solutions of Maxwell's equations
(or a specialization thereof):  
\begin{subequations}
  \label{eq:maxwell}
  \begin{align}
    \nabla\times\mbE + \frac{1}{c}\pderiv{\mbB}{t} &= 0, \label{eq:faraday}\\
    c^2\nabla\times\mbB - \pderiv{\mbE}{t} &= \frac{\mbj}{\epsilon_0},
    \label{eq:ampere}\\ 
    \nabla\cdot\mbE &= \frac{\rho}{\epsilon_0}, \label{eq:gauss}\\ 
    \nabla\cdot\mbB &= 0,\label{eq:mag_guass}
  \end{align}
\end{subequations}
where $\epsilon_0$ is the permittivity of free space and $c$ is the
\emph{in vacuo} speed of light.  The total charge density, $\rho$, and
the total current, $\mbj$, are the sums over contributions
from all species $\alpha$,
\begin{subequations}
  \label{eq:dens}
  \begin{align}
    \rho(\mbx,t) &= \sum\limits_{\alpha} \rho_{\alpha}(\mbx,t) = 
    \sum\limits_{\alpha}
    q_{\alpha} \int_{\field{R}^d} f_{\alpha}\ d\mbv,\\
    \mbj(\mbx,t) &= \sum\limits_{\alpha}\mbj_{\alpha}(\mbx,t) 
    = \sum\limits_{\alpha} q_{\alpha} \int_{\field{R}^d} \mbv
    f_{\alpha}\ d\mbv,
  \end{align}
\end{subequations}
and these moments of the distribution function nonlinearly couple
Maxwell's equations to the Vlasov equation. 

The Vlasov-Maxwell system and related models are fundamental
non-equilibrium descriptions of plasma dynamics.  Non-equilibrium
kinetic effects in plasmas play a crucial role in fusion applications.
Understanding and controlling wave-particle interactions 
is important to the success of inertial confinement fusion, where resulting
resonant responses can interfere with the intended deposition of laser
energy~\cite{LiAmBeGlGlHaKaLaSu03}.  In magnetic confinement fusion,
gyrokinetic models, which are a reduced form of the Vlasov
equations~\cite{FrCh82,Ha88,DiLoDu92}, 
are used to better understand the physical mechanisms controlling the
core conditions, in particular micro-turbulence, which is at the origin
of the so-called anomalous transport~\cite{TaCh05}.  

The Vlasov model also has applicability beyond fusion plasmas.
Collisionless shocks in astrophysics, which are thought to be driven by
electrostatic and electromagnetic
instabilities~\cite{weibel59,sagdeev66,tidman71}, can be accurately 
modeled by the Vlasov-Maxwell system.  The Vlasov-Poisson system, where
Gauss' Law~\eqref{eq:gauss} is sufficient to describe the relationship
between the electrostatic field and the charge density, is being used
in particle beam accelerator design~\cite{chombopic03}.  Laser isotope
separation is another application area for Vlasov-Maxwell
models~\cite{FrWa02}. 
 
While these kinetic models may have great relevance, their numerical 
approximation for problems of interest have been constrained primarily
by computational cost.  For \mbox{$N=3$}, the distribution functions 
in the full Vlasov model have a phase-space domain of six dimensions.
Directly discretizing phase
space, an approach alternatively referred to as \emph{grid-based},
\emph{Eulerian}, or \emph{continuum} methods, incurs a computational cost
that scales geometrically with the number of dimensions. 
Thus, while for over forty years work has been done on the continuum
numerical discretization of the Vlasov
equation~\cite{ShKn74,ChKn76,Kl87,BeGhJoShFiFe90,BeGhBe99,Fi99,NaYa99,FiSoBe01,NaTaYaTa01,ArVa02,BeSo03,BrGrSaVaVi04,GuHaPaSo04,BrVa05,HaLaGu05,KlFa94,WaSuSa01,FiSo03,PoShKa05,SircN09,sonnendrucker},
continuum Vlasov methods have been applied primarily to
lower-dimensional -- so-called 1D+1V and 1D+2V -- problems.  Application
of continuum Vlasov to four dimensions (2D+2V) and above has been
limited~\cite{SircN09,BaBeBrCoHi11,sonnendrucker,strozzi12}. 
In contrast, the \emph{particle-based} particle-in-cell
(PIC)~\cite{BiLa1991} method has dominated kinetic Vlasov simulation.
PIC methods use Monte-Carlo sampling techniques in velocity space to
reduce the high-dimensional cost and evolve ``clouds'' of particles through a
Lagrangian form of the Vlasov equation.  Maxwell's equations, however,
are solved on an overlaid \new{computational} mesh (hence, ``in cell'').
While this approach is generally less expensive than continuum Vlasov
discretization, PIC results contain inherent statistical 
noise that generally vanishes only as the square root of the number of particles.

As computer speed and memory have increased, direct discretization of
the Vlasov-Maxwell system has become more feasible, but hardware
improvements alone are insufficient to make full-phase-space, continuum
Vlasov codes practical.  However, as with PIC methods, tremendous savings
could be realized if continuum approaches could reduce the number of
cells used to represent phase space.  One means to this end is to employ
adaptive mesh refinement and to resolve only those regions of phase
space of greatest variation or importance. For instance, block-structured
adaptive mesh refinement (AMR) in phase space could
concentrate cells in the vicinity of localized structure, such as
particle trapping regions.  In addition, flux-based 
explicit Eulerian schemes have time-step restrictions that, for Vlasov,
are typically dominated by the maximum particle velocity limits of the
phase-space domain.  AMR allows for high-aspect ratio cells in these
regions, which can result in significant increases in time step size 
without a loss of accuracy, since little particle density or variation is
present at these extreme velocity boundaries.

Adaptive mesh refinement has a limited history in Vlasov simulation.
AMR has been used with PIC methods in the simulation of heavy-ion
accelerators~\cite{Va02b,chombopic03}. Recent work in this area uses
a wavelet-based approach~\cite{GuHaPaSo04,HaLaGu05}, where 
the semi-Lagrangian interpolation is based upon a multi-level wavelet
basis and where the local depth of the wavelet hierarchy is used to
increase or decrease the local mesh refinement.  This approach
generates a near-optimal grid, but progress in this direction seems to
have stalled.  It may be the case that the less regular grid structure
may introduce other complications, for example, in the construction of
moments, that make this approach less attractive.

In this paper, we present a block-structured adaptive mesh refinement
approach suitable for the continuum discretization of the
Vlasov-Maxwell system.  As a proof-of-concept, we demonstrate the ideas
and techniques in the context of a simpler system, the Vlasov-Poisson
model, which is presented in Section~\ref{sec:model} along with the basic
flux-based Eulerian discretization we employ.  Thus, we will not
address the control of electromagnetic wave reflections at coarse-fine
interfaces; methods to minimize such reflections are addressed elsewhere
in the literature~\cite{Va01,Va02b}.  
In Section~\ref{sec:bsamr}, we discuss the block-structured AMR
strategy, its benefits, and the challenges presented by Vlasov
problems.  We specifically address a subset of these issues that we have
resolved in order to demonstrate a successful Vlasov-AMR implementation.
Sample calculations are presented in Section~\ref{sec:results}, and we
conclude with a discussion of the future algorithmic advances that will
allow additional gains from AMR applied to Vlasov simulation.

\section{Model Problem and Discretization}
\label{sec:model}
Both for our purposes here as well as for many physically interesting problems,
the Vlasov-Maxwell system can be significantly simplified by assuming an
electrostatic limit with stationary ions. The electrostatic limit
corresponds to an assumption of small magnetic field strength. The
assumption of stationary ions is appropriate when the ion time scales
are large compared to that of the electrons, which is typically the
case. With these assumptions, the Vlasov equation~\eqref{eq:vlasov_nc} for the
electron probability density function $f$ becomes
\begin{equation}
  \pt{\f}+\vv\cdot\nabla_{\xv}\f-\Ev\cdot\nabla_{\vv}\f=0,
  \label{eq:vlasov_electron_3d}
\end{equation}
under a suitable nondimensionalization.  Here $\Ev$ is the electric field,
$\xv$ is the physical space and $\vv$ is the velocity. In the electrostatic limit,
only Gauss' law~\eqref{eq:gauss} is relevant.  Representing the electric
field in terms of an electrostatic potential, $\Ev = \nabla_{\xv} \phi$,
Gauss' law becomes the Poisson equation:
\begin{equation}
\nabla_{\xv}^2\phi  = \rho_e - \rho_i = \int \f \,d\vv-1.
  \label{eq:poisson2d}
\end{equation}
Here the constant, unit background charge, $\rho_i=1$, is the
manifestation of the immobile ion (proton) assumption.   

For the purposes of discussing the new adaptive discretization
algorithms, we make one final simplifying assumption of a 
so-called 1D+1V phase space (\ie, one spatial dimension and
one velocity dimension).  The final system of governing equations is
thus succinctly written as
\begin{subequations}
\begin{gather}
  \pt{\f}+v\px{\f}+\px{\phi}\pv{\f}  = 0, \label{eq:vlasov1d}\\
  \pxx {\phi} = \int_{-\infty}^{\infty}f\,dv-1.\label{eq:poisson1d}
\end{gather}
\end{subequations}
Note that, as characteristic of all Vlasov-Maxwell-type systems, we have
a higher-dimensional variable, $f(x,v)$, coupled to a lower-dimensional
variable, $\phi(x)$.

In order to discretize the model Vlasov-Poisson system, we restrict our
attention to a finite domain. For the physical coordinate we let
$x\in[-L,L]$ and apply periodic boundary conditions. Other boundary
conditions are also possible, but for the initial-value problems
considered here, a periodic condition is appropriate. For the velocity
coordinate, we truncate the domain and consider
$v\in[v_{\hbox{min}},v_{\hbox{max}}]$. This introduces an artificial
boundary where we apply a characteristic boundary condition. Outgoing
characteristics are extrapolated and incoming characteristics carry
values from an unperturbed Maxwellian distribution. 

Our discretization follows the Eulerian finite-volume formulation developed 
in~\cite{CoDoHiMcMa09,CoDoHiMa10}. The 1D Vlasov
equation~\eqref{eq:vlasov1d} is rewritten in flux-divergence form as  
\begin{equation}
  \dt{\f}+\dx{\left(v\f\right)}-\dv{\left(E\f\right)}=0.
  \label{eq:vlasov1ddiv}
\end{equation}
Phase space is divided into cells using a Cartesian grid with mesh
spacings $\Delta x$ and $\Delta v$ in the $x$- and $v$-dimensions
respectively.  Integrating over a single computational cell and
dividing by the volume $\Delta x\Delta v$, we obtain the exact system of
ordinary differential equations
\begin{equation}
     \frac{d}{dt}\bar{f}_{ij} = -\frac{1}{\Delta x}\left(
    \langle v\f\rangle_{i+\half,j}-\langle v\f\rangle_{i-\half,j}\right)
  \quad+\frac{1}{\Delta v}\left(
    \langle E\f\rangle_{i,j+\half}-\langle E\f\rangle_{i,j-\half}\right), 
  \label{eq:semi-discrete}
\end{equation}
where the cell average $\bar{f}_{ij}$ is defined as 
\begin{equation*}
  \bar{f}_{ij} \equiv \frac{1}{\Delta x\Delta v}\int_{V_{ij}}
  f dx dv.
\end{equation*}
As in~\cite{CoDoHiMcMa09,CoDoHiMa10}, the angle bracket notation is used to
indicate face averages, for example,
\begin{equation*}
  \langle \f\rangle_{i+\half,j} = \frac{1}{\Delta v}
  \int_{v_{j-1/2}}^{v_{j+1/2}} \f(x_{i+1/2},v) dv.
\end{equation*}
The face-averaged fluxes are approximated to fourth order as
\begin{align*}
  \langle v\f\rangle_{i+\half,j} &\approx
  \bar{v}_{j} \langle \f\rangle_{i+\half,j} 
  +\frac{\Delta v}{24} \left( \langle \f\rangle_{i+\half,j+1}-\langle \f\rangle_{i+\half,j-1}\right),\\
\begin{split}
  \langle E\f\rangle_{i,j+\half} &\approx
  \bar{E}_{i} \langle \f\rangle_{i,j+\half} 
  -\frac{1}{48}
   \left(\bar{E}_{i+1}-\bar{E}_{i-1}\right)
   \left(\langle f\rangle_{i+1,j+\half}-\langle f\rangle_{i-1,j+\half}\right).
\end{split}
\end{align*}
Notice that, because $v$ is only a function of $v$ and $E$ is only a
function of $x$, the notion of a face average is redundant, and the angle
bracket is replaced by an an overbar. For more details concerning this
high-order finite-volume formalism, refer to~\cite{CoDoHiMcMa09,CoDoHiMa10}.

The quantity $\bar{v}_j$ is directly computed as an exact cell average
(recall that $v$ is an independent variable). The cell-averaged
electric field is computed from a potential $\phi$; to fourth-order, this is
\new{\begin{equation*}
\bar{E}_i \approx \frac{1}{12\Delta
  x}\left[8(\bar{\phi}_{i+1}-\bar{\phi}_{i-1})-\bar{\phi}_{i+2}+\bar{\phi}_{i-2}\right]. 
\end{equation*}}
The cell-averaged potential is obtained by solving a discretization of the
Poisson equation~\eqref{eq:poisson1d} :
\begin{equation}
\label{eq:discrete_potential}
30\bar{\phi}_{i}-16(\bar{\phi}_{i+1}+\bar{\phi}_{i-1})+(\bar{\phi}_{i+2}+\bar{\phi}_{i-2}) = 12\Delta x\bar{\rho}_i,
\end{equation}
where 
\begin{align*}
  \bar{\rho}_i = 1-\Delta v\sum_{j=-v_{\text{max}}}^{v_{\text{max}}} \bar{f}_{ij}.
\end{align*}
This discretization leads to a linear system with a nearly pentadiagonal
matrix (boundary conditions slightly alter the pentadiagonal structure).

For reasons explained in Section~\ref{sec:bsamr}, the Poisson problem is
always represented on the finest mesh in 
configuration space, and so the resulting linear algebra problem can be
LU-decomposed once for each level of refinement and stored. Periodic
boundary conditions in $x$ lead to a singular system, which is a
well-known problem that is easily addressed by projecting out the portion of
$\bar{\rho}(x)$ residing in the null space of the matrix.  This amounts
to ensuring that $\sum_i\bar{\rho}(x_i)=0$, and in so doing, we ensure
that $\bar{\phi}(x)$ is normalized around zero.  Of course, since we
take a derivative of $\bar{\phi}(x)$ to get  
$\bar{E}(x)$, the offset has no effect on the solution.

To complete the description of the discretization, a procedure to derive
face averages from cell averages must be identified.  We use the scheme
developed in~\cite{BaHi10,BaBeBrCoHi11}, which has the property that, for
well-represented solutions, a fourth-order centered
approximation is used. As solution features become sharp on a given
mesh, upwind numerical dissipation is introduced to smooth out those
features consistently. The scheme is described in detail
in~\cite{BaHi10,BaBeBrCoHi11}, but we provide a brief overview here as well.

We focus on the determination of the face average $\langle
\f\rangle_{i+\half,j}$; other averages follow similar derivations. The
scheme has many similarities to 
the popular WENO~\cite{Shu97} method and uses many of the tools developed 
in the literature on that topic. The face average is constructed as a weighted
sum of two third order approximations: 
\begin{equation}
  \langle \f \rangle_{i+\half,j} \approx {w}_{i+\half,j,L}\langle \f \rangle_{i+\half,j,L}+{w}_{i+\half,j,R}\langle \f \rangle_{i+\half,j,R},
  \label{eq:limitedFace}
\end{equation}
with
\begin{equation}
  \langle \f \rangle_{i+\half,j,L} \approx \frac{1}{6}\left(-\bar{f}_{i-1,j}+5\bar{f}_{i,j}+2\bar{f}_{i+1,j}\right)
  \label{eq:thirdLeft}
\end{equation}
and
\begin{equation}
  \langle \f \rangle_{i+\half,j,R} \approx \frac{1}{6}\left(2\bar{f}_{i,j}+5\bar{f}_{i+1,j}-\bar{f}_{i+2,j}\right).
  \label{eq:thirdRight}
\end{equation}
Here the ``L'' and ``R'' indicate left- and right-biased, third-order
approximations. 
With ideal weighting, ${w}_{i+\half,j,L}={w}_{i+\half,j,R}=\half$,
equation~\eqref{eq:limitedFace} becomes the centered, fourth-order
approximation. Using the standard WENO methodology, provisional weights,
$\hat{w}_{i+\half,j,L}$ and $\hat{w}_{i+\half,j,R}$, are determined. 
To maximize the upwind diffusion in the final numerical method, we
assign the larger weight to the upwind, third-order approximation and 
the smaller weight for the downwind, third-order stencil.  Thus the final
weights are determined as
\begin{equation}
\begin{array}{cl}
  \hbox{if } \left(v_{j} > 0\right), \hspace{0.1in} & \left\{
  \begin{array}{lcl}
  w_{i+\half,j,L} & = & \max( \hat{w}_{i+\half,j,L},\hat{w}_{i+\half,j,R}),\smallskip \\
  w_{i+\half,j,R} & = & \min( \hat{w}_{i+\half,j,L},\hat{w}_{i+\half,j,R}),
  \end{array}
  \right. \medskip \\
  \hbox{else }&
  \left\{
  \begin{array}{lcl}
  w_{i+\half,j,L} & = & \min( \hat{w}_{i+\half,j,L},\hat{w}_{i+\half,j,R}),\smallskip \\
  w_{i+\half,j,R} & = & \max( \hat{w}_{i+\half,j,L},\hat{w}_{i+\half,j,R}).
  \end{array}
  \right.
\end{array}
\end{equation}
Note that, as with traditional WENO schemes, convergence rates near
certain types of  critical points (points with many zero derivatives)
may be less than optimal.  Additional modifications to the provisional
weights can be made to alleviate this deficiency~\cite{henrick05}.  

For the temporal discretization of the semi-discrete Vlasov
equation~\eqref{eq:semi-discrete}, any stable method can be used.  We
choose the standard explicit fourth-order Runge-Kutta scheme.  At each
stage in the Runge-Kutta update, we solve the discrete potential
equation~(\ref{eq:discrete_potential}) prior to evaluating the
phase-space flux divergence as given by the right-hand side
of~\eqref{eq:semi-discrete}.

Consider the ODE initial value problem
\begin{subequations}
  \begin{gather}
  \deriv{f}{t} = L(f,t),\\
  f(0) = f_0.
  \end{gather}
\end{subequations}
The RK4 discretization for the ODE between time level $n$ and $n+1$ is
\begin{subequations}
  \label{eq:rk4}
  \begin{gather}
    f^{n+1} = f^n + \Delta t \sum_{s=1}^4 b_sk_s,\\
    k_s = L\left(f^{(s)},t^n+c_s\Delta t\right),\\
    f^{(s)} = f^n + \alpha_s\Delta t k_{s-1},
  \end{gather}
\end{subequations}
with $\balpha = [0,1/2,1/2,1]$, $\mbb = [1/6,1/3,1/3,1/6]$, and $\mbc = [0,1/2,1/2,1]$.
Acknowledging that the operator $L$ is, in our case, of flux-divergence
form, we can write, for example, in one dimension, 
\begin{subequations}
  \label{eq:flux_accum}
  \begin{align}
    f^{n+1}_i &= f^n_i + \Delta t \sum_{s=1}^4 b_sk_{i,s},\\
    &= f^n - \Delta t \sum_{s=1}^4 b_s\left[F_{i+1/2}\left(f^{(s)}\right)-F_{i-1/2}\left(f^{(s)}\right)\right],\\
    &= f^n - \Delta t \left[\sum_{s=1}^4 b_sF_{i+1/2}\left(f^{(s)}\right)-\sum_{s=1}^4 b_sF_{i-1/2}\left(f^{(s)}\right)\right],\\
    &= f^n - \Delta t \left[F^*_{i+1/2}-F^*_{i-1/2}\right],
  \end{align}
\end{subequations}
where $F^*_{i+1/2}$ are accumulated interface fluxes.


\section{Block Structured AMR Algorithms}
\label{sec:bsamr}
Block-structured adaptive mesh refinement~\cite{BeOl84,BeCo89} is a
natural fit for certain Vlasov-Maxwell problems.  Frequently, important
fine-scale features in phase space, which could substantially benefit
from higher resolution, only occupy limited regions in phase space. 
\begin{figure}[t]
  \centering
  \setlength{\unitlength}{0.01in}
  \begin{picture}(600,270)(0,0)
  \put(150,265){\large$\mathcal{G}_C$}
  \put(400,265){\large$\mathcal{G}_H$}
  \put(75,25){\small$i$}
  \put(87,25){\small$0$}
  \put(105,25){\small$1$}
  \put(123,25){\small$2$}
  \put(142,25){\small$3$}
  \put(160,25){\small$4$}
  \put(179,25){\small$5$}
  \put(197,25){\small$6$}
  \put(216,25){\small$7$}
  \put(62,31){\small$j$}
  \put(62,45){\small$0$}
  \put(62,64){\small$1$}
  \put(62,82){\small$2$}
  \put(62,100){\small$3$}
  \put(62,119){\small$4$}
  \put(62,138){\small$5$}
  \put(62,157){\small$6$}
  \put(62,175){\small$7$}
  \put(62,194){\small$8$}
  \put(62,213){\small$9$}
  \put(58,230.5){\small$10$}
  \put(150,0){$x$}
  \put(35,130){$v$}
  \put(80,40){\includegraphics*[width=1.5in]{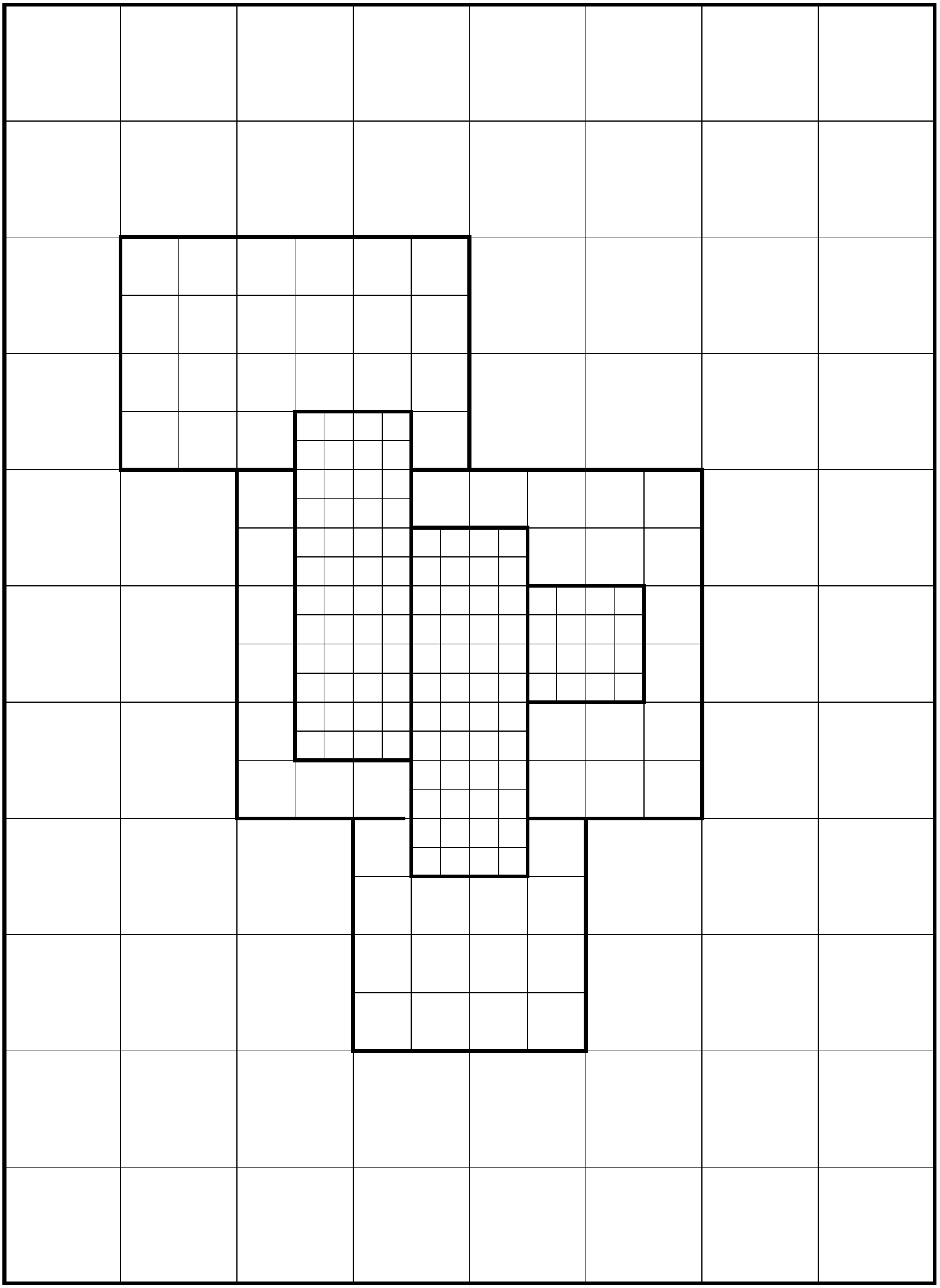}}
  \put(300,40){\includegraphics*[width=2.5in]{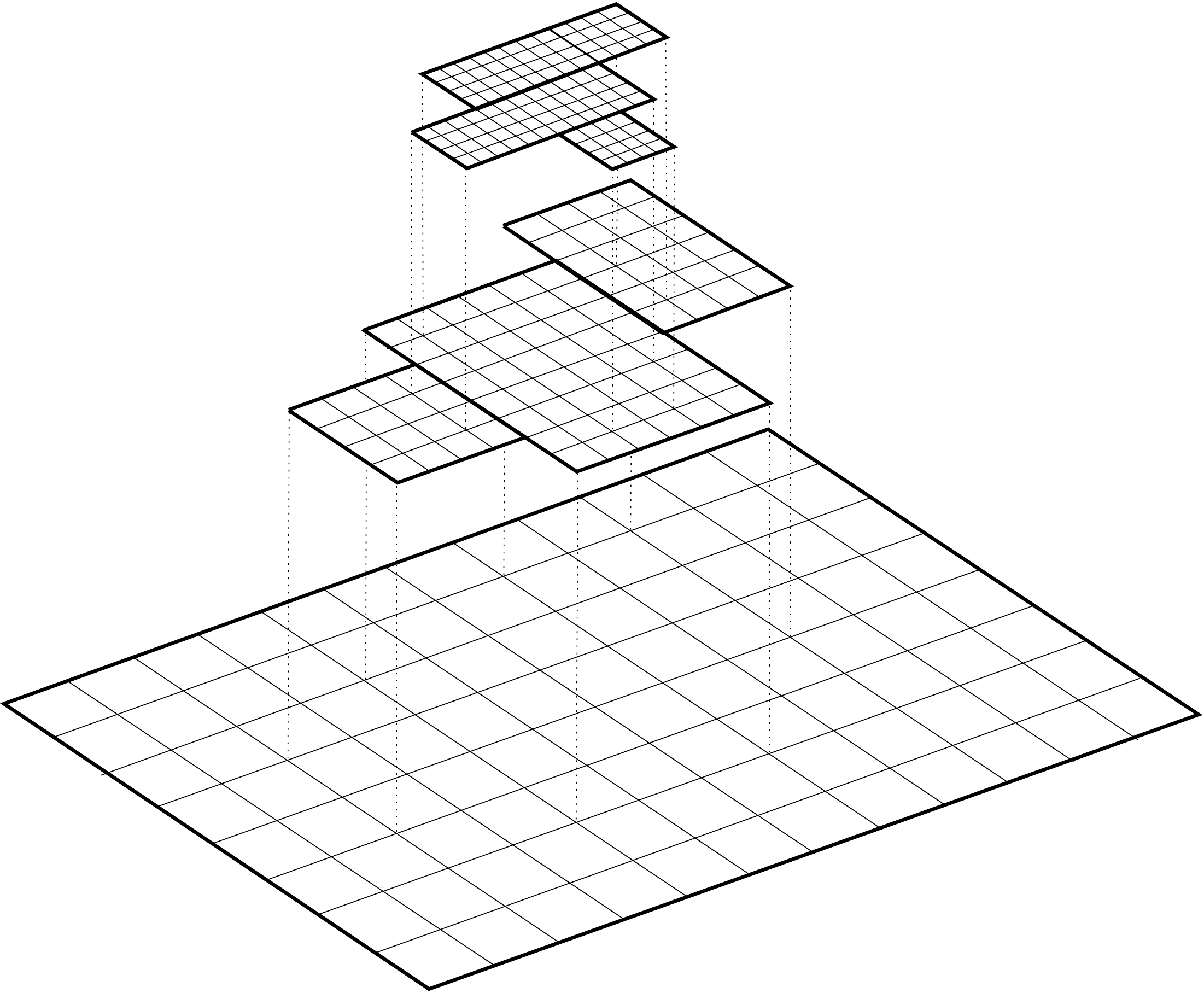}}
  \end{picture}
  \caption{An example of a three-level, block-structured AMR hierarchy.
    On the left, the composite refined grid $\mathcal{G}_C$ is shown.
  On the right, the corresponding mesh hierarchy $\mathcal{G}_H$ with
  overlapping patches is shown.
  All patches on the same level have the same refinement ratio relative
  to the coarsest level; in this case, the refinement ratios
  are two and four for the intermediate and finest levels,
  respectively.  Note that each level is comprised of a collection of
  patches completely contained within the patches of the next coarser
  level.}
  \label{fig:amrgrid}
\end{figure}
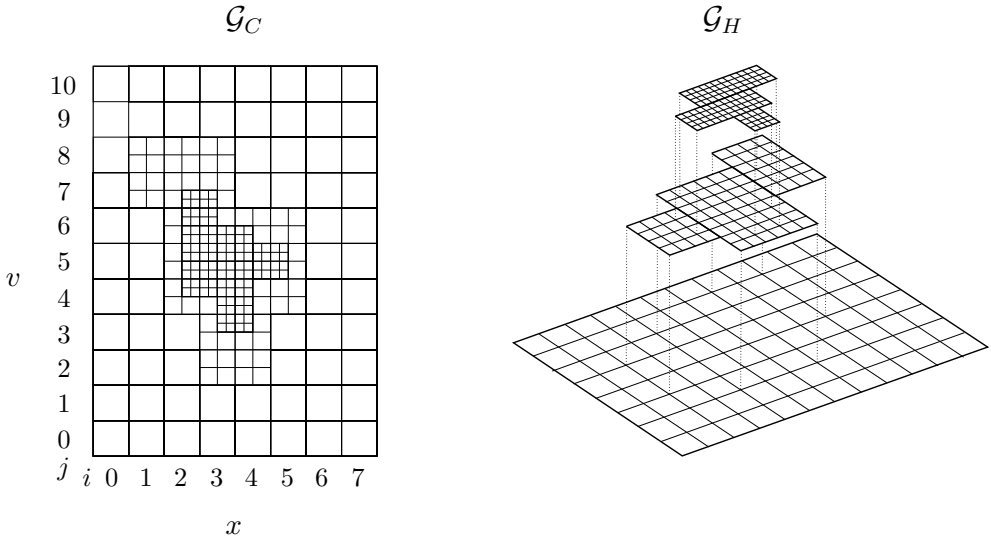
In contrast to the semi-structured, octree-based grids that were
used in the earlier Vlasov-AMR work~\cite{BeSo03}, hierarchical
block-structured AMR is  
based upon rectangular grid patches at different refinement levels in a
global Cartesian index space, as shown in Figure~\ref{fig:amrgrid}.
Using a local error estimate or some  
detection of rapid variation in the solution to identify 
regions of interest, cells that should be refined are tagged.  Tagged
cells are grouped and expanded minimally to form rectangular patches
that are inserted into the next level in the hierarchy.  Slightly larger
refinement regions can be used to reduce the frequency of regridding.
The refinement process can be repeated recursively to form a hierarchy of
refinement levels, each composed of multiple patches.

Connectivity information is kept to a minimum in this scheme because 
the global Cartesian index space provides a simple mechanism by which to
identify the relationships between patches and levels.  Within a level,
patches contiguous in indices are neighboring, and across levels, the
same is true, after adjusting by the net refinement ratio between the
two levels.  In general, for explicit methods, communication between
patches is accomplished through ghost cells.  As an additional savings,
by maintaining a consistent solution on all patches, even those covered
by patches on a finer level, time refinement algorithms that allow for
nested subcycling on finer levels can be devised.

Despite all of the previous work on block-structured AMR, applying the
technique to Vlasov simulation introduces several new challenges.  First
and foremost, at least two mesh hierarchies must be maintained: one in
the $\field{R}^N$ configuration space and one in the
$\field{R}^N\times\field{R}^M$ phase space.  Different kinetic
species will, in general, have different masses and temperatures;
the bulk of the corresponding particles will therefore occupy different
ranges of particle velocity, and the structures arising from resonant
responses will occur in different regions of phase space. Thus, each
kinetic species should have its own mesh hierarchy. Thus, new algorithms
for the simultaneous advancement and \new{coordination} of multiple
hierarchies are required, and more importantly, efficient algorithms to
enable communication between the hierarchies are required.  
From a parallel implementation perspective, a hierarchy for each species
also allows for increased task parallelism when each hierarchy is
assigned to a subset of processors; with no collisions, kinetic species
only communicate through the lower-dimensional configuration space, so,
given the electromagnetic state, high-dimensional flux computations and 
updates can naturally be done in parallel.

In this paper, it is our goal to demonstrate solutions to the
fundamental issues that must be addressed to make effective use of AMR
in Vlasov simulation.  Specifically, we will discuss:
\begin{itemize}
\item \textbf{Basic modifications due to discretization.}  Using
  high-order finite volume and a high-order multi-stage schemes departs
  somewhat from the standard, nominally second-order block-structured AMR
  approach.  We describe the modified algorithms we use, for example, the
  intra-hierarchy interpolation operations and the synchronous time
  integration algorithm.
\item \textbf{Inter-hierarchy transfer operations.}  The coupling of
  problems of different dimension and their representation on separate
  hierarchies necessitates the creation of inter-hierarchy
  reduction and injection transfer algorithms.  We discuss algorithms
  that achieve this efficiently. 
\item \textbf{\new{Regridding for multiple AMR hierarchies.}}
  Regridding hierarchies, when multiple related hierarchies are present,
  requires additional constructs for coordination.
\end{itemize}
In the following subsections, we address each of these areas, describing
in more details the issues involved and explaining our solution
approach.  We will not specifically address efficient parallel decomposition
strategies in this work.

These new AMR algorithms, combined with the high-order discretizations
presented in Section~\ref{sec:model} have been implemented in the Vlasov code
\Valhalla\footnote{Vlasov Adaptive Limited High-order Algorithms for
  Laser Applications}.  This code makes use of the 
block-structured AMR library SAMRAI~\cite{HoKo02}, which has 
been used in prior plasma-fluid simulations~\cite{Dorr02}.  
SAMRAI is capable of handling dimensions above three as well as the
simultaneous existence of multiple and lower hierarchies, possibly of
different dimension.  A graph-based, distributed implementation of mesh
metadata is employed within SAMRAI to provide excellent scaling to tens
of thousands of processors.  
SAMRAI also provides fairly sophisticated, high-level
AMR algorithm abstractions, but the simultaneous advancement of multiple
related hierarchies, as required by Vlasov-Poisson, does not fit into
these integration strategies and has thus required substantial
additional development. 

\subsection{Basic modifications due to discretization}
Our base discretization uses a method-of-lines approach, where
spatial operators are first discretized using a nominally fourth-order
spatial discretization and then the resulting semi-discrete system is
integrated using the standard four-stage, fourth-order explicit
Runge-Kutta method.  
Fortunately, for a high-order finite-volume implementation,
the restriction algorithm to obtain a coarse cell average from fine cell
averages remains the simple summation used for lower-order schemes.

\subsubsection{Synchronous, multi-stage time advancement algorithm}
In practice, an asynchronous process with time step subcycling on finer
cells is typically used for explicit, space-time
discretizations~\cite{BeOl84}, but we chose to start with a synchronous
update for simplicity.  
For a synchronized update (\ie, a single $\Delta t$ for all levels), 
the RK4 algorithm~\eqref{eq:rk4}-\eqref{eq:flux_accum} for a
conservation law on a single-hierarchy is summarized in
Algorithm~\ref{alg:ssthu}. 
Looping over stages, the predictor states and fluxes are computed, and
the fluxes are accumulated.  The predictor-state algorithm is laid out
in Algorithm~\ref{alg:mspsc},
and the flux divergence and accumulation algorithm is sketched in
Algorithm~\ref{alg:msrhsfa}.  
We note that, for this conservative form, we accumulate a flux
variable as in~\eqref{eq:flux_accum} so that we can construct a flux
divergence using a temporally fourth-order flux for the final update.
Such flux accumulation eliminates an explicit re-fluxing step, since the
final update can be done 
from finest to coarsest levels, and the accumulated flux can be averaged
down so that a single update using the highest-quality flux can be done
on each level.  The update is computed from the
accumulated fluxes as shown in Algorithm~\ref{alg:msfduc}, and
regridding is done if a user-defined number of time steps have elapsed.
\begin{algorithm}[H]
\caption{Multi-Level, Single-Hierarchy Flux-Divergence RK4 Advance}
\label{alg:ssthu}
\begin{algorithmic}
  \State $k\gets0$
  \ForAll{Stages $s \gets 1,4$}
  \State \textsc{ComputePredictorState}($k,s,f_{pred}$)
  \State \textsc{ComputeRHS}($f_{pred},s,k,F_{accum}$)
  \EndFor
  \State \textsc{ComputeUpdate}($F_{accum},f_{new}$)
  \If{time to regrid}
  \State Regrid all levels
  \EndIf
  \State Compute next $\Delta t$
\end{algorithmic}
\end{algorithm}
\begin{algorithm}[H]
\caption{Multi-Stage Predictor State Computation}
\label{alg:mspsc}
\begin{algorithmic}
  \Procedure{ComputePredictorState}{$k,s,f_{pred}$}
  \State $t \gets t_{old} + c_s\cdot \Delta t$
  \State $f_{pred} \gets f_{old} + \alpha_{s}\cdot \Delta
  t\cdot  k$
  \State Interpolate up to ghost cells on finer levels of $f_{pred}$
  \State Exchange ghost cells on each level of $f_{pred}$
  \State Apply boundary conditions to $f_{pred}$
  \EndProcedure
\end{algorithmic}
\end{algorithm}
\begin{algorithm}[H]
\caption{Multi-Stage Right-Hand Side Evaluation and Flux Accumulation}
\label{alg:msrhsfa}
\begin{algorithmic}
  \Procedure{ComputeRHS}{$f_{pred},s,k,F_{accum}$}
  \ForAll{Levels $l\gets 1,L$}
  \ForAll{Patches $p$}
  \State $F_{pred}\gets$ \textsc{computeFluxes}$(f_{pred},t)$
  \EndFor
  \State Exchange fluxes between patches on level $l$
  \EndFor
  \ForAll{Levels $l \gets L,1$ }
  \ForAll{Patches $p$}
  \State $k \gets$ \textsc{fluxDivergence}$(F_{pred})$
  \State $F_{accum}\gets F_{accum} + b_s \cdot F_{pred}$
  \EndFor
  \EndFor
  \EndProcedure
\end{algorithmic}
\end{algorithm}
\begin{algorithm}[H]
\caption{Multi-Stage, Multi-Level Flux-Divergence Update Computation}
\label{alg:msfduc}
\begin{algorithmic}
  \Procedure{ComputeUpdate}{$F_{accum},f_{new}$}
  \ForAll{Levels $l \gets L,1$ }
  \ForAll{Patches $p$}
  \State $\delta f \gets \textsc{fluxDivergence}(F_{accum})$
  \State $f_{new} \gets f_{old} + \Delta t \cdot \delta f$
  \State Coarsen fluxes down to level $l-1$
  \EndFor
  \EndFor
  \State Coarsen fine data down for $f_{new}$
  \EndProcedure
\end{algorithmic}
\end{algorithm}

To integrate the Vlasov-Poisson system, the time advancement algorithm
must be adapted to allow the simultaneous advancement of multiple
phase-space hierarchies.  In addition, the Poisson equation represents
an 
instantaneous constraint, and we chose most self-consistent strategy of
re-evaluating the Poisson equation at each predictor state.  An alternative 
possibility is to extrapolate $\phi$ in time to avoid some
of the intermediate field solves and the associated parallel
synchronization; investigating this approach is left for future work.

\begin{algorithm}[h]
\caption{Synchronous, Multi-Stage, Vlasov-Poisson Multi-Hierarchy Advance}
\label{alg:smsvphu}
\begin{algorithmic}
  \State $k\gets0$
  \ForAll{Stages $s \gets 1,4$}
  \ForAll{Hierarchies $H$}
  \State \textsc{ComputePredictorState}($k,s,f_{pred}$)
  \EndFor
  \State \textsc{ComputeInstantaneousConstraints}($f_{pred},\phi$)
  \ForAll{Hierarchies $H$}
  \State \textsc{ComputeRHS}($f_{pred},\phi,s,k,F_{accum}$)
  \EndFor
  \EndFor
  \ForAll{Hierarchies $H$}
  \State \textsc{ComputeUpdate}($F_{accum},f_{new}$)
  \EndFor
  \If{time to regrid}
  \State Regrid all hierarchies
  \EndIf
  \State \textsc{ComputeInstantaneousConstraints}($f_{new},\phi$)
  \State Compute next $\Delta t$
\end{algorithmic}
\end{algorithm}
The associated modifications to Algorithm~\ref{alg:ssthu} are shown in
Algorithm~\ref{alg:smsvphu}.  The main differences are that the major
steps are each now computed for all hierarchies and that additional
steps to evaluate instantaneous constraints have been inserted on
predicted or updated states are obtained.  Note that we do not recompute
the potential until after any possible regridding since the regridding
step for the configuration space hierarchy is not independent of the
phase space hierarchies in our current implementation.  More details
about this are given in Section~\ref{sec:regrid}.

\subsubsection{Conservative, limited, high-order interpolation algorithm}
\label{sec:prolong}
Fine-patch cells are filled from coarse cells either when new fine-level
patches are created or when fine-patch ghost cells at coarse-fine
interfaces are filled.  For second-order discretizations of first-order 
differential operators, slope-limited linear
reconstruction is generally used to obtain fine-cell averages from the
coarse grid while controlling non-physical oscillations.  To obtain 
a fourth-order reconstruction while controlling oscillations, 
several techniques exist, including least squares~\cite{McCo10},
unfiltered~\cite{Barad2007}, and explicitly-filtered~\cite{Ray2007}
high-order interpolations.  We adopt a slightly
different approach and make use of standard WENO5~\cite{Shu97}
interpolants that have been analytically integrated to obtain explicit
cell-average interpolation formulas.  

We assume cell-averaged data $\bar{u}_{\mbi}$ on a coarse mesh with mesh
size $\mbh$ and an overlapping fine mesh with mesh size $\mbh^f$, such
that 
\begin{equation}
  h^f_j = h_j/R_j,\qquad j=1,2,\ldots,D,
\end{equation}
where each $R_j$ is a positive integer.  Our goal is to construct a
high-order approximation to fine-mesh cell-averaged values
$\bar{u}^f_{\mbi_f}$ such that the integral over the fine mesh exactly
equals the integral over the coarse mesh.  In addition, since
initialization of fine mesh from coarse mesh may be done in regions of
high gradients, we seek an adaptive interpolation scheme that will
inhibit the creation of unphysical oscillations.

\begin{figure}[t]
  \centering
  \includegraphics*[width=5in]{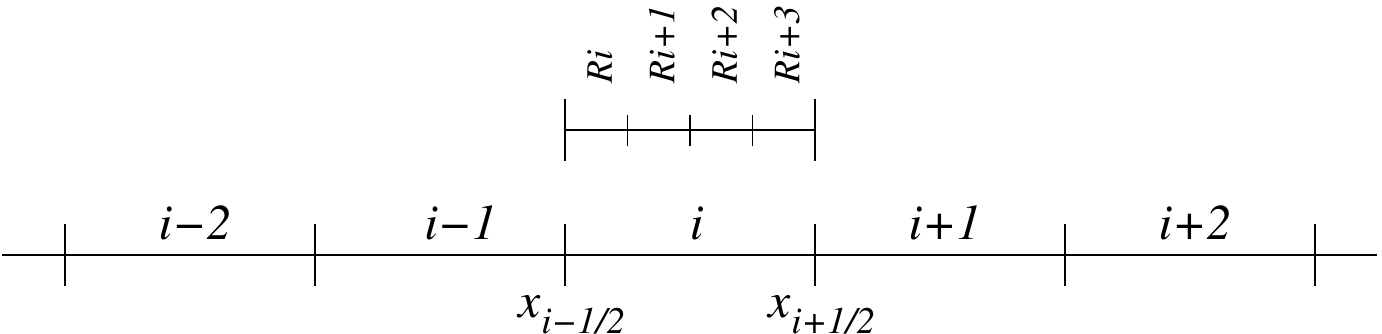}
  \caption{Relationship of fine to coarse mesh in global index space
  with a refinement ratio of $R=4$.  The five coarse cells shown are
  used to determine the cell averages in the four fine cells that
  subdivided cell $i$.}  
  \label{refineop1d}
\end{figure}
For our fourth-order discretization, the five-point WENO5 scheme is sufficient.  In the
general approach to obtain an interpolation in cell $i$, one is given
five cell-averages $\bar{u}_{i+e}$, $e=-2,-1,0,1,2$,
and a location $x_{i-1/2}\le x\le x_{i+1/2}$, as shown in
Figure~\ref{refineop1d}.  It is useful at this 
point to define some auxiliary quantities:
\begin{subequations}
  \label{aux1}
  \begin{align}
    D_{i+n}&=\bar{u}_{i+n+1}-\bar{u}_{i+n},\qquad n=-2,-1,0,1,\\
    \Delta_{i+p}&=D_{i+p}-D_{i+p-1},\qquad p=-1,0,1.
  \end{align}
\end{subequations}
The algorithm proceeds for a uniform mesh as follows:
\begin{enumerate}
  \item Compute smoothness detectors $\beta^{(r)}_i$, $r=0,1,2$:
    \begin{subequations}
      \begin{align}
        \beta^{(0)}_i &= \frac{13}{12}\Delta_{i+1}^2+\frac{1}{4}\left(D_{i+1}-3D_{i}\right)^2,\\
        \beta^{(1)}_i &= \frac{13}{12}\Delta_{i}^2+\frac{1}{4}\left(D_{i}+D_{i-1}\right)^2,\\
        \beta^{(2)}_i &= \frac{13}{12}\Delta_{i-1}^2+\frac{1}{4}\left(3D_{i-1}-D_{i-2}\right)^2;
      \end{align}
    \end{subequations}
  \item Compute the absolute interpolation weights $\alpha^{(r)}_i$, $r=0,1,2$:
    \begin{equation}
      \alpha^{(r)}_i = d_r/(\epsilon+\beta^{(r)}_i)^2,
    \end{equation}
    where $d_0=3/10$, $d_1=3/5$, $d_2=1/10$, and $\epsilon$ is a small positive
    value to avoid division by zero (typically $\epsilon=10^{-6}$);
  \item Compute the relative interpolation weights $\omega^{(r)}_i$, $r=0,1,2$:
    \begin{equation}
      \omega^{(r)}_i = \alpha^{(r)}_i/\left(\D\sum\limits_{s=0}^{2}\alpha^{(s)}_i\right);
    \end{equation}
  \item Compute the interpolants $v^{(r)}_i(x)$, $r=0,1,2$:
    \begin{equation}
      v^{(r)}_i(x)=h\sum\limits_{m=1}^2 \left[\left(\sum\limits_{j=0}^{m-1}
      \bar{u}_{i+j-r}\right)
      \left(
      \frac{\sum\limits_{\substack{l=0\\l\ne m}}^2\left[\prod\limits_{\substack{q=0\\q\ne m,l}}^2\left(x-x_{i-1/2}-h(q-r)\right)\right]}{\prod\limits_{\substack{l=0\\l\ne m}}^2h(m-l)}
      \right)
      \right];
      \label{interp_r}
    \end{equation}
  \item Compute the combined interpolant $v_i(x)$:
    \begin{equation}
      v_i(x) = \sum\limits_{r=0}^{2}\omega^{(r)}_i v^{(r)}_i(x).
    \end{equation}
\end{enumerate}
The result $v_i(x)$ is an interpolant constructed from cell average values
such that 
\begin{equation}
  \int\limits_{x_{i-1/2}}^{x_{i+1/2}} v_i(x)\ dx = h\bar{u}_i.
\end{equation}
In smooth regions, $v_i(x)$ is an $\Order{h^5}$ approximation pointwise; in regions
where under-resolution generates oscillations, the scheme drops to
$\Order{h^3}$ pointwise (at worst) by adapting its stencil through the nonlinear
weights so as to bias towards interpolations that are less oscillatory.

For adaptive mesh refinement, we can analytically
integrate~\eqref{interp_r} over the fine-mesh cells to arrive at
simple algebraic equations for the fine-mesh cell-averages.  For a
refinement ratio of $R$, we integrate~\eqref{interp_r} over each of the
intervals $[x_{i-1/2}+sh/R,x_{i-1/2}+(s+1)h/R]$, $s=0,1,\new{\ldots},R-1$, (see
Figure~\ref{refineop1d}).  Define for $r=0,1,2$, $A^{(r)}_i$,
$B^{(r)}_i$, and $C^{(r)}_i$:  
\begin{equation}
  \label{aux2}
  \begin{array}{rl}
    A^{(0)}_i &= \bar{u}_i + \frac{1}{6}(2D_{\new{i}+1}-5D_{\new{i}}),\\
    A^{(1)}_i &= \bar{u}_i - \frac{1}{6}(D_{\new{i}}+2D_{\new{i}-1}),\\
    A^{(2)}_i &= \bar{u}_i - \frac{1}{6}(4D_{\new{i}-1}-D_{\new{i}-2}),\\
  \end{array}\qquad
  \begin{array}{rl}
    B^{(0)}_i &= 2D_{\new{i}}-D_{\new{i}-1},\\
    B^{(1)}_i &= B^{(2)}_{\new{i}} = D_{\new{i}-1},\\
  \end{array}\qquad
  \begin{array}{rl}
    C^{(0)}_i &= \Delta_{\new{i}+1},\\
    C^{(1)}_i &= \Delta_{\new{i}},\\
    C^{(2)}_i &= \Delta_{\new{i}-1}.
  \end{array}
\end{equation} 
Then the three fine-mesh cell-averaged interpolated values in cell $(Ri+s)$ are
\begin{equation}
  \left(\bar{u}^f_{Ri+s}\right)^{(r)} = A^{(r)}_i + B^{(r)}_i \left(\frac{2s+1}{2R}\right) + C^{(r)}_i
  \left(\frac{3s^2+3s+1}{6R^2}\right),
  \label{interp_r_avg}
\end{equation}
for $r=0,1,2$.  Note that, \new{to ensure exact conservation to
  round-off, we renormalize the average of the fine cells to the
  original coarse cell average by
\begin{equation}
  \left(\bar{u}^f_{R(i+1)-1}\right)^{(r)}_{\text{renorm}} =
  \left(\bar{u}^f_{R(i+1)-1}\right)^{(r)} + \bar{u}_i -
  \sum\limits_{s=0}^{R-1} \left(\bar{u}^f_{Ri+s}\right)^{(r)}/R;
\end{equation}
this ensures that the truncation errors are equidistributed amoung the
sub-cells.}

In implementation, advantage can be made of the many repeated factors.
Notably, for any given coarse cell $i$, the fifteen auxiliary
variables~\eqref{aux1} and~\eqref{aux2} and the $\omega^{(r)}_i$ only need be
computed once for the $R$ fine cells in cell $i$.  Similarly, for a
fixed refinement $R$, the $2(R-2)$ functions of $s=0,1,\cdots,R-2$
in~\eqref{interp_r_avg} are the same for any coarse cell $i$.

\begin{figure}[t]
  \centering
  (a)\includegraphics*[height=1.75in]{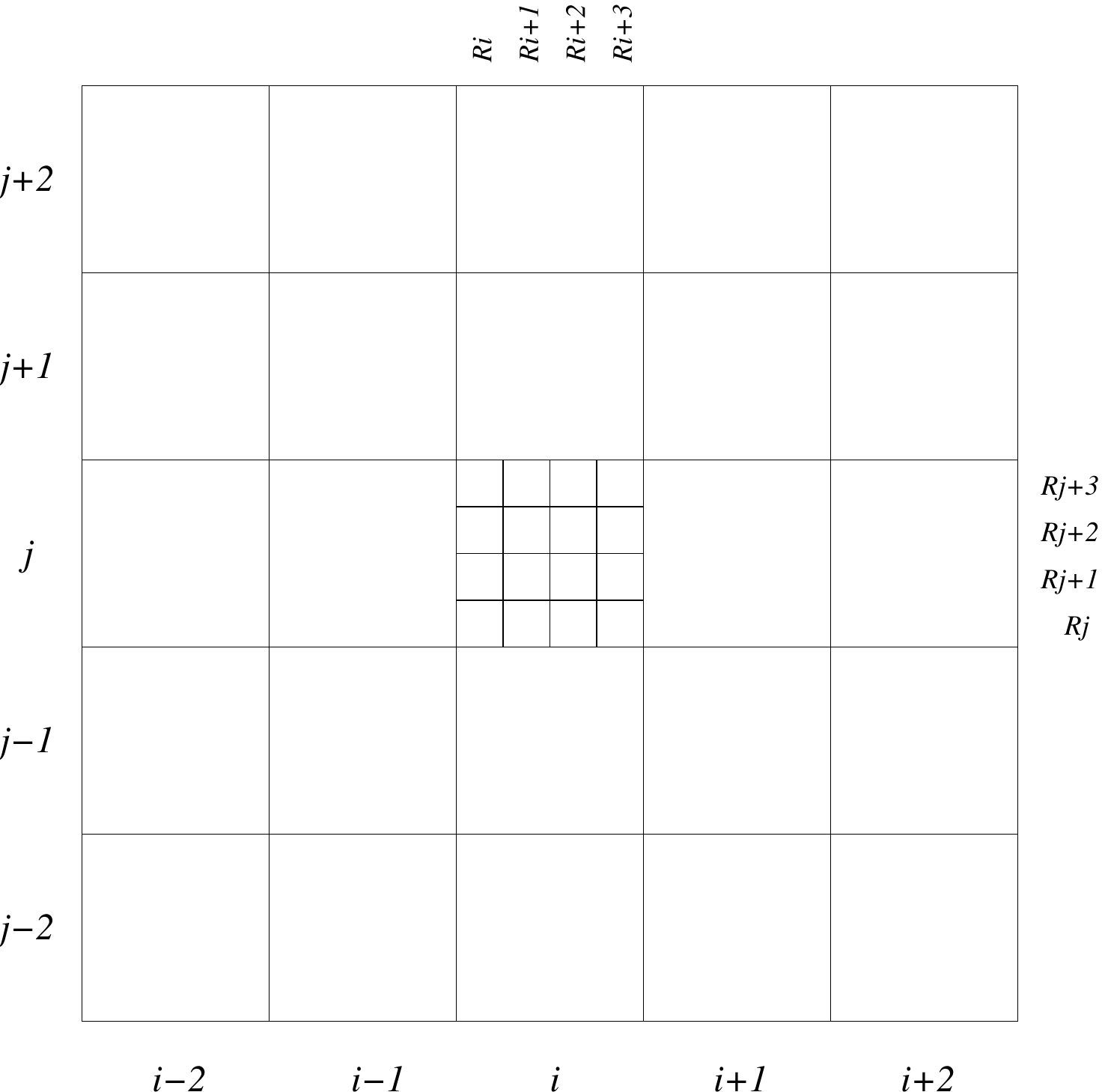}\hfill
  (b)\includegraphics*[height=1.75in]{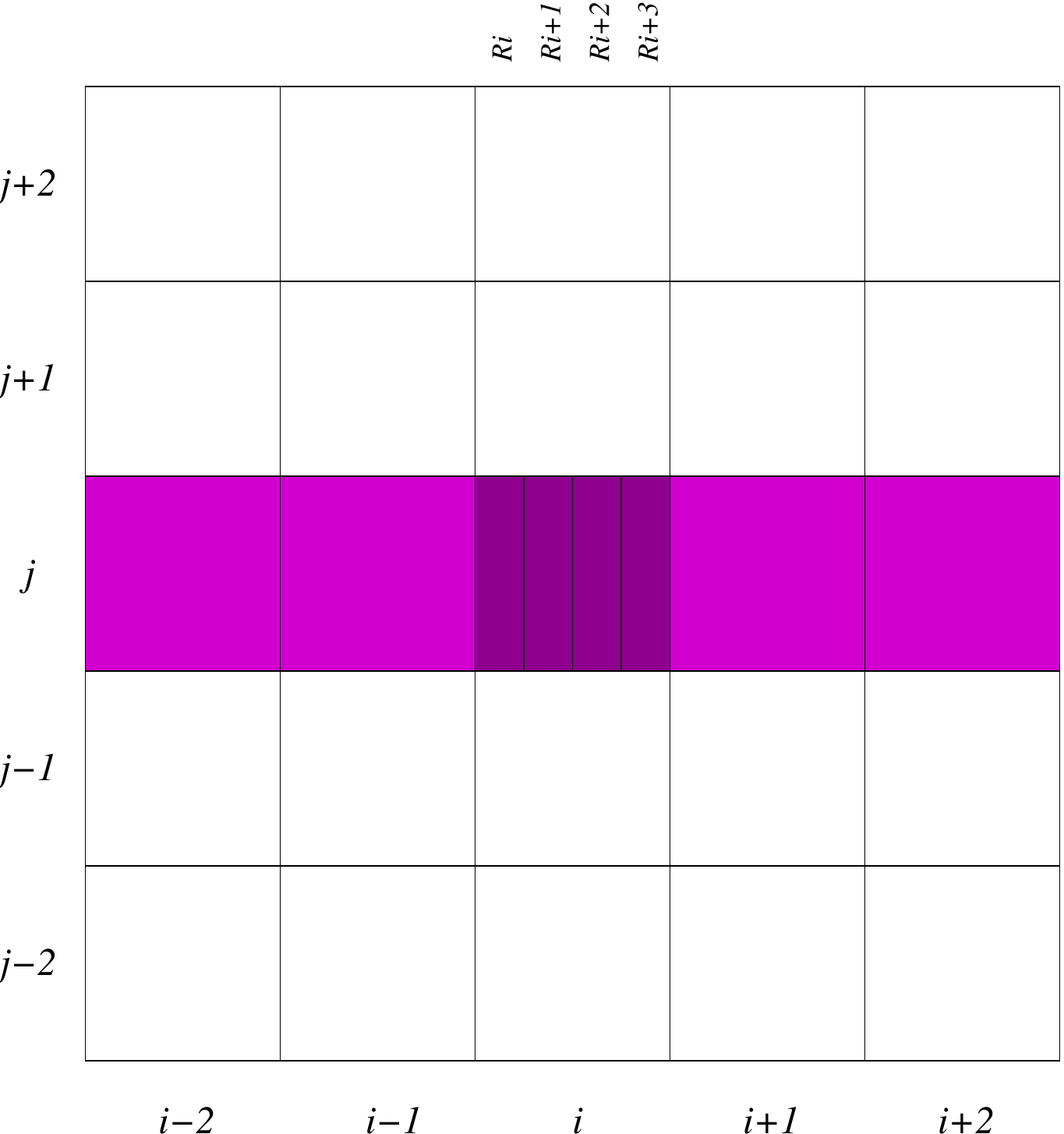}\hfill
  (c)\includegraphics*[height=1.75in]{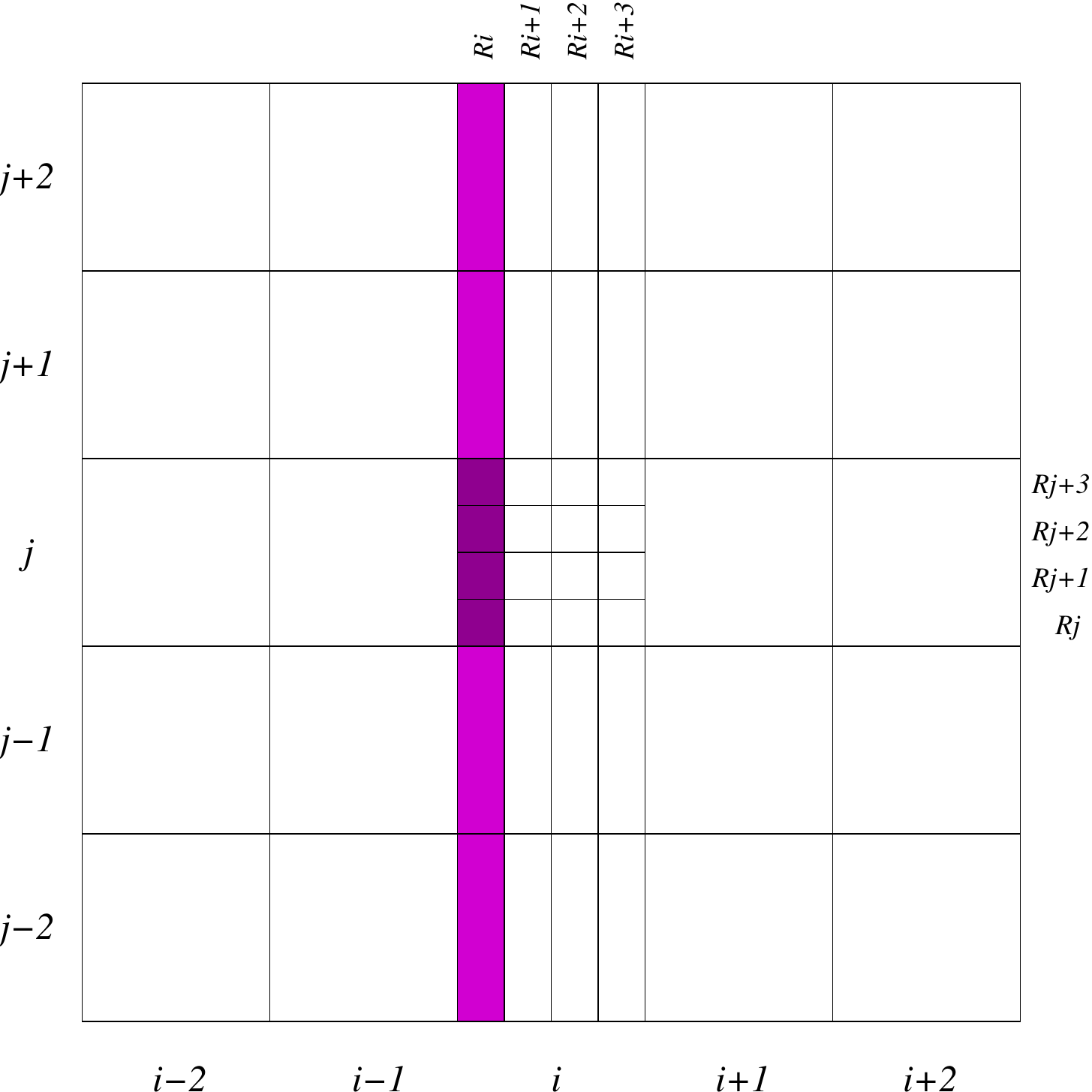}
  \caption{(a) The twenty-five coarse cells in 2D used to interpolate the
  sixteen fine-cell averages in cell $i$ for a uniform refinement of
  $R=4$. (b) The cells involved in the partial interpolation in the
  $x$-direction for cell $i$.  The result are the four cell averages that are
  fine in the $x$-direction but coarse in the $y$-direction. (c)
  The cells involved in the partial interpolation in the
  $y$-direction for sub-cells at fine-grid location $Ri$.  This is
  repeated for all fine-grid locations in the $x$-direction.}
  \label{refineop2d}
\end{figure}
The direct, though not most efficient, extension of the one-dimensional
algorithm to multiple dimensions is to apply the method
dimension-by-dimension.  Thus, it is sufficient to build code to handle
the 1D 
problem, and the multi-dimensionality is handled through data
management, \ie, the input provided to the 1D routines and the memory
destinations to which the results are written.

Consider the 2D case where the refinement ratios are $R_0$ and
$R_1$ in the $x_0$- and $x_1$-directions, respectively.  An example with
$R_0=R_1=R=4$ is shown in Figure~\ref{refineop2d}(a).  In cell $i$, we
first compute cell averages for cells refined only in $x_0$
using~\eqref{interp_r_avg}.  This is shown in Figure~\ref{refineop2d}(b).
The result for each cell $i$ is $R_0$ new sub-cell values.

The same operation is then applied in $x_1$-direction, but the input
values are no longer the coarse-grid averages, but are now the
partially-refined averages from the previous step.  This is shown in
Figure~\ref{refineop2d}(c).  The result 
for each fine cell $Ri+s$ is $R_1$ sub-cell values, and since there are
$R_0$ $x_1$-interpolations per coarse cell $i$, $R_0R_1$ sub-cell values.

\subsection{Inter-hierarchy transfer operations}
\label{sec:moment}
Data transfer between hierarchies of different dimensionality requires
the formulation of special algorithms and auxiliary data structures.
While the injection of lower-dimensional data in the higher-dimensional
space is a straight-forward constant continuation in the new dimensions,
the reduction of higher-dimensional data into the lower-dimensional
space requires the application of an operator across the dimensions that
are removed, such as the integrals in the moment reductions~\eqref{eq:dens}.  

The application of reductions across an AMR hierarchy is not in itself
new.  For example, \new{the computation of mathematical norms is} frequently executed on AMR
hierarchies, and any norm is the reduction of higher-dimensional data
into a scalar value.  Lower-dimensional slices across a
hierarchy are often used for visualization.  A special case
of such a slice reduction was developed for the laser-plasma interaction code
ALPS~\cite{Dorr02}, where the paraxial light wave sweeps required
plasma densities on lower-dimensional planar slices.
The challenge for the Vlasov system is that the reductions
are the result of accumulation.  Spatial fidelity and accuracy must be
maintained while 
accumulating across the velocity dimensions, and such reductions must be done
efficiently and without double-counting (recall the overlapping mesh
hierarchy shown in Figure~\ref{fig:amrgrid}).

In addition to the need to obtain data at equivalent resolution, the act
of orchestrating a reduction operation across a hierarchy in parallel
requires several auxiliary structures.  In fact, to preserve generality
in the mesh refinement, auxiliary data structures are also helpful
for injection operations.  We next discuss two moment reduction
algorithms that have been developed for the \Valhalla code, followed by
a brief description of the associated injection algorithm.

\subsubsection{Moment reduction algorithm}
\begin{figure}[t]
  \centering
  \setlength{\unitlength}{0.01in}
  \begin{picture}(600,50)(0,0)
  \put(70,25){\small$i$}
  \put(105,25){\small$0$}
  \put(152,25){\small$1$}
  \put(202,25){\small$2$}
  \put(252,25){\small$3$}
  \put(302,25){\small$4$}
  \put(352,25){\small$5$}
  \put(402,25){\small$6$}
  \put(452,25){\small$7$}
  \put(300,0){$x$}
  \put(80,40){\includegraphics*[width=4in]{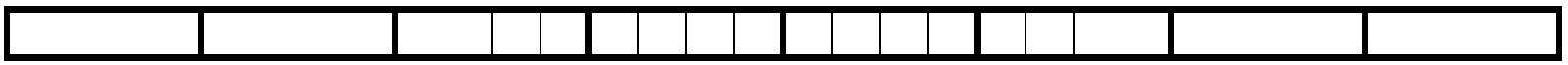}}
  \end{picture}
  \caption{The configuration-space composite grid corresponding to the
  composite grid $\mathcal{G}_C$ depicted in Figure~\ref{fig:amrgrid}
    after reduction.}
  \label{fig:compreducedgrid}
\end{figure}
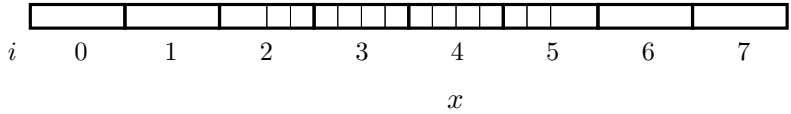
Consider the composite grid depicted in Figure~\ref{fig:amrgrid}.  Let us assume
that we will accumulate on coarse-grid index $j$ for coarse-grid index
$i=1$.  Along $i=1$ there are cells of two resolutions since there are
two cells, $j=7,8$, that have been refined.  If we do not preserve the
finest resolution in the accumulation on $j$, we will lose some
known information about the spatial structure in the remaining
$x$-direction.  To preserve the finest-resolution spatial information,
we should subdivide in the $x$-direction to obtain cells of uniform resolution
(in $i$).  Using this principle, the corresponding composite grid after
reduction is shown in Figure~\ref{fig:compreducedgrid}.

One might consider subdividing coarse cells without spatially
reconstructing the data within the coarse cell, but this would result in
an $\Order{h}$ error in the summation.  To see this, consider that, for
a refinement ratio of $R$, the relationship between a fine grid cell
average and a coarse grid average in a single dimension is
\begin{equation}
  \bar{u}^f_{Ri+s}= \bar{u}_{i}-h\left(\frac{R-2s-1}{2R}\right)\left.\pd_xu\right|_{i+\frac{2s+1}{2R}}+\Order{h^2},
\end{equation}
where $s=0,1,\ldots,R-1$.  Thus, to preserve higher-order accuracy, all
coarse data must be reconstructed before averaging to the finest 
resolution and before applying the reduction operation.

\new{One can make use of the limited interpolation operators defined in
Section~\ref{sec:prolong}.  However, the data to be reduced should be
well-resolved (or else it would have been refined), so a less expensive
option is to use a linear interpolation.  One can construct such an
interpolant by averaging over sub-cells the fifth-order interpolant,
\begin{equation}
v(x)=\sum\limits_{j=-2}^2\gamma_j\!\left(\frac{x}{h}\right)\bar{u}_{i+j}+\Order{h^5},
\end{equation}
with
\begin{subequations}
\begin{align}
\gamma_{-2}(\eta) &= \left(5\eta^4-20\eta^3+15\eta^2+10\eta-6\right)/120,\\
\gamma_{-1}(\eta) &= \left(-20\eta^4+60\eta^3+30\eta^2-150\eta+54\right)/120,\\
\gamma_{0}(\eta) &= \left(30\eta^4-60\eta^3-120\eta^2+150\eta+94\right)/120,\\
\gamma_{1}(\eta) &= \left(-20\eta^4+40\eta^3+90\eta^2-10\eta-26\right)/120,\\
\gamma_{2}(\eta) &= \left(5\eta^4-15\eta^2+4\right)/120,
\end{align}
\end{subequations}
as was done to arrive at~\eqref{interp_r_avg}.  The resuling formula for the
fine-mesh interpolations is 
\begin{equation}
\bar{u}^f_{Ri+s} = \sum\limits_{j=-2}^2b_j(s)\bar{u}_{i+j}+\Order{h^5}
\end{equation}
with
\begin{subequations}
\begin{align}
b_{-2}(s) &= \left(p_4(s)-5p_3(s)+5p_2(s)+5p_1(s)-6\right)/120,\\
b_{-1}(s) &= \left(-4p_4(s)+15p_3(s)+10p_2(s)-75p_1(s)+54\right)/120,\\
b_{0}(s) &= \left(6p_4(s)-15p_3(s)-40p_2(s)+75p_1(s)+94\right)/120,\\
b_{1}(s) &= \left(-4p_4(s)+5p_3(s)+30p_2(s)-5p_1(s)-26\right)/120,\\
b_{2}(s) &= \left(p_4(s)-5p_2(s)+4\right)/120,
\end{align}
\end{subequations}
where
\begin{equation}
p_k(s) = \sum\limits_{j=0}^k\frac{k!}{j!(k-j)!}s^j.
\end{equation}
}

\newcommand{\restr}{\operatorname{restr}}
\newcommand{\repl}{\operatorname{repl}}
At this point, it is helpful to introduce some notation.  We denote an
$N$-vector of integers by
$\mbi=(i_0,i_1,\ldots,i_{N-1})\in\field{Z}^{N}$ and a patch, 
$\mathcal{P}$, by a pair of $N$-vectors that indicate
the lower and upper cells of 
the patch: $\mathcal{P}=[\mbi_{\text{lo}},\mbi_{\text{hi}}]$.  The
restriction from an 
$N$-vector to a $(N-1)$-vector by removing the $j$-th element is
$\restr_j$, \eg,
\begin{equation}
  \restr_j\mbi=(i_0,i_1,\ldots,i_{j-1},i_{j+1},\ldots,i_{N-1})\in\field{Z}^{N-1}.
\end{equation}
and we define $\restr_j\mathcal{P}=[\restr_j\mbi_{\text{lo}},\restr_j\mbi_{\text{hi}}]$.
We also define a replacement operator $\repl_j({a,b})$ that operates on
patches and that replaces the $j$-th element of the lower and upper
$N$-vectors by $a$ and $b$, respectively:
\begin{equation}
  \repl_j({a,b})\mathcal{P}=[(i_0,i_1,\ldots,i_{j-1},a,i_{j+1},\ldots,i_{N-1}),(i_0,i_1,\ldots,i_{j-1},b,i_{j+1},\ldots,i_{N-1})].
\end{equation}
A collection of patches at a refinement level $l$ is denoted by
$\mathcal{L}_l$, and a patch hierarchy $\mathcal{G}_H$ is defined as the
set of refinement levels with an $N$-vector refinement ratio defined
between each level and the coarsest ($l=\new{0}$), \ie, 
\begin{equation}
  \mathcal{G}_H =\{\mathcal{L}_l, \new{0}\le l\le \new{L-1}: \exists\ \mbr^{\new{l+1}}_{\new{l}}\in\field{Z}^N,\ \new{0}\le l\le L-\new{2}\}.
\end{equation}
The directional refinement and coarsening operators,
$\mathcal{R}_j^{a,b}$ and $\mathcal{C}_j^{a,b}$, respectively, refine
and coarsen the $j$-th direction of a patch by the ratio defined between
refinement levels $a$ and $b$, that is,
\begin{subequations}
\begin{align}
\mathcal{R}_j^{a,b}\mathcal{P} &= \repl_j({R_j^{a,b}i_j,R_j^{a,b}(i_j+1)-1})\mathcal{P},\\
\mathcal{C}_j^{a,b}\mathcal{P} &= \repl_j({C_j^{a,b}i_j,\lfloor{C_j^{a,b}(i_j+1)-1}\rfloor})\mathcal{P},
\end{align}
\end{subequations}
where
$R_j^{a,b}=\prod_{l=0}^{\new{b-1}}(\mbr^{\new{l+1}}_{\new{l}})_j/\prod_{l=0}^{\new{a-1}}(\mbr^{\new{l+1}}_{\new{l}})_j$
and $C_j^{a,b}=1/R_j^{a,b}$.

Again referring to Figure~\ref{fig:amrgrid}, we wish to compute the
reduction on the composite grid $\mathcal{G}_C$, but in fact we have the
hierarchy grid $\mathcal{G}_H$.  The standard technique for computing a
reduction across a hierarchy without double counting is to use a mask to
zero out the contributions from coarse grid regions that are overlapped
by fine grid.
\new{L}et \new{$P_l$}
be the number of \new{patches} in \new{level}
$\mathcal{L}_l$; let $\mathcal{I}_j^{a,b}$ be the 
the interpolation operator in direction $j$ that refines the local data from
the resolution of level $a$ to that of level $b$; and let $\mu_{\mbi}^p$
be the masking operator on patch $\mathcal{P}_p$ that sets the data in
cell $\mbi$ 
to zero if the cell is covered by a patch at a finer level.  We further
assume, without loss of generality, that the phase-space integer vector
$\mbi$ is ordered in such a 
way so that \new{elements 0 through $N-1$ correspond to configuration-space 
indices, and elements $N$ through $N+M-1$ correspond to velocity space indices.}
%

The reduction operation to construct the charge density is
\begin{equation}
\begin{split}
  \rho_{\mbi_c} &=
 1-V_v\sum\limits_{d=N}^{N+M-1}\sum\limits_{j_d=0}^{J_d-1}f_{\mbi} = 1-V_v
 \sum\limits_{d=N}^{N+M-1}\sum\limits_{l=\new{0}}^{\new{L-1}}\sum\limits_{p=\new{0}}^{\new{P_l-1}}\sum\limits_{j_d=j^p_{d,\text{lo}}}^{j^p_{d,\text{hi}}}\mu_{\mbi}^p\prod\limits_{d'=0}^{N-1}\mathcal{I}_{j_{d'}}^{l,\new{L-1}}f^{l}_{\mbi},\\
&= 1-V_v
 \sum\limits_{l=\new{0}}^{\new{L-1}}\sum\limits_{p=\new{0}}^{\new{P_l-1}}\sum\limits_{d=N}^{N+M-1}\sum\limits_{j_d=j^p_{d,\text{lo}}}^{j^p_{d,\text{hi}}}\mu_{\mbi}^p\prod\limits_{d'=0}^{N-1}\mathcal{I}_{j_{d'}}^{l,\new{L-1}}f^{l}_{\mbi},
\end{split}
\end{equation}
where $\mbi_c=\prod_{d=N}^{N+M-1}\restr_d\mbi\in\field{Z}^N$ is the
configuration space index and where $V_v=\prod_{d=N}^{N+M-1}h_d$.
In words, the distribution function on each patch $p$ is first refined up to
the finest level in the configuration space directions, then masked, and
then accumulated.  Note that the mask operator and the
interpolation operator do not commute; after summing over the velocity
-space indices, $j_d$, with a mask, the resulting partial sums would in
general have discontinuities at mask boundaries.  Thus, using a masking
procedure, one must reconstruct in the higher-dimensional space, which is
more expensive that reconstructing in the lower-dimensional space.

To facilitate the inter-dimensional communication, two intermediate data
structures are used.   The \emph{partial reduction level}
is a level of \emph{overlapping} patches at the finest resolution in
configuration space, where
the patches are projections of every patch in the
phase-space hierarchy:
\begin{equation}
  \mathcal{L}_{\text{partial}}=\left\{\prod\limits_{d'=0}^{N-1}\mathcal{R}_{d'}^{l,\new{L-1}}\prod\limits_{d=N}^{N+m-1}\restr_d\mathcal{P},\quad\forall\ \mathcal{P}\in\mathcal{G}_H\right\}.
\end{equation}
The \emph{total reduction level}, $\mathcal{L}_{\text{total}}$,
\emph{disjoint} covering of patches of the configuration space domain
that, aside from the resolution, is independent of all hierarchies.
These structures are created given a phase-space hierarchy
$\mathcal{G}_H$ and configuration space-hierarchy $\mathcal{G}^c_H$
and exist until regridding occurs.

\begin{figure}[t]
  \centering
  \includegraphics*[width=4.5in]{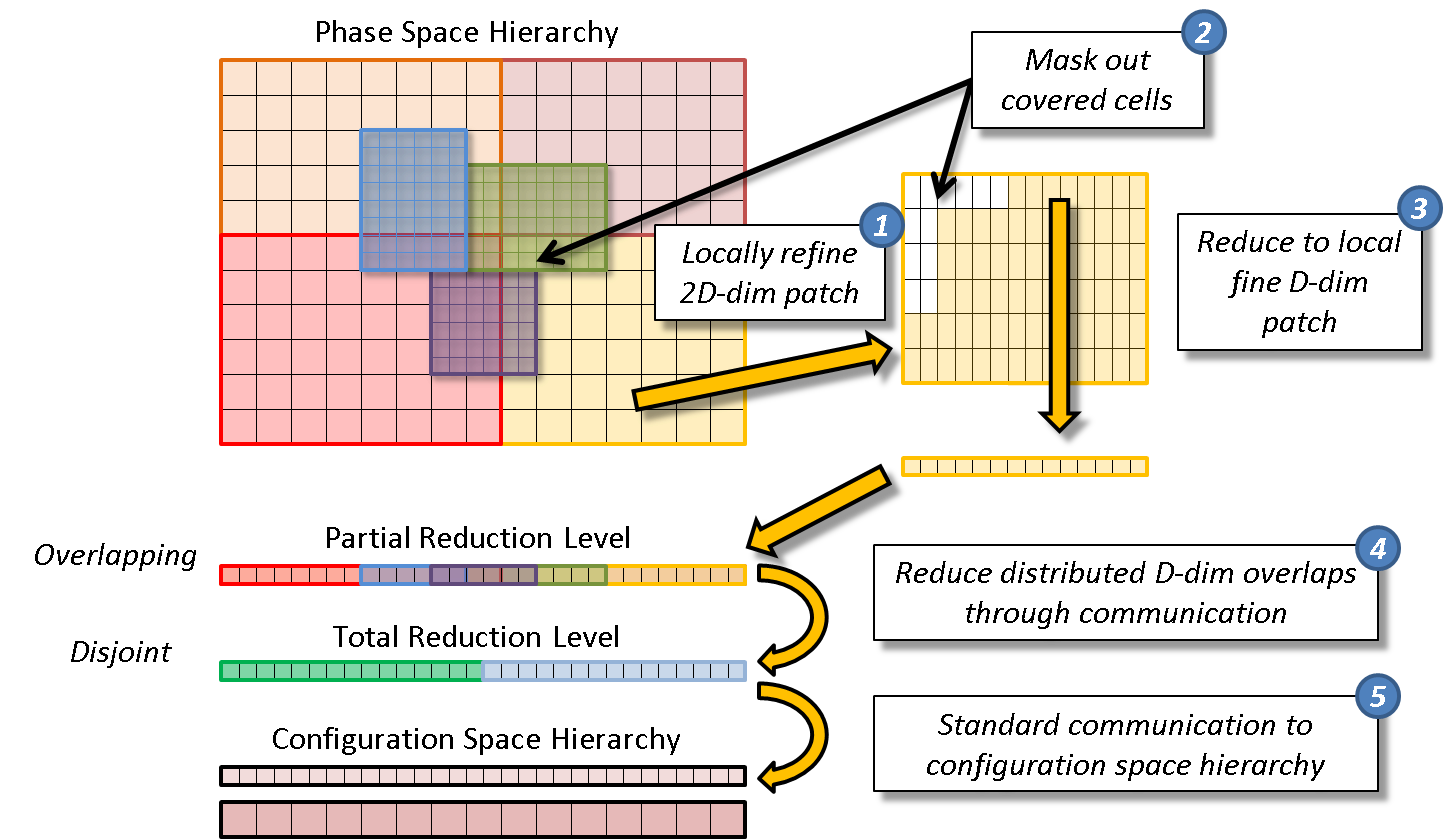}
  \caption{Graphical depiction of the Mask Reduction Algorithm.}
  \label{fig:reduction}
\end{figure}
\begin{algorithm}[H]
\caption{Mask Reduction}
\label{alg:mreduce}
\begin{algorithmic}
  \ForAll{levels $\mathcal{L}\in\mathcal{G}_H$}
  \ForAll{patches $\mathcal{P}\in\mathcal{L}$}
  \State Refine data in non-reduction directions
  \State Mask out \new{covered cells}
  \State Find patch $\mathcal{P}_{\text{p}}(\mathcal{P})\in\mathcal{L}_{\text{partial}}$ corresponding to patch $\mathcal{P}$
  \State Sum reduce to partial sum patch $\mathcal{P}_{\text{p}}$
  \EndFor
  \EndFor
  \ForAll{patches $\mathcal{P}_{\text{t}}\in\mathcal{L}_{\text{total}}$}
  \State Accumulate data from co-located patches $\mathcal{P}_{\text{p}}$
  \EndFor
  \ForAll{levels $\mathcal{L}\in\mathcal{G}^c_H$}
  \ForAll{patches $\mathcal{P}\in \mathcal{L}$}
  \State Copy data from co-located patches $\mathcal{P}_{\text{t}}$
  \EndFor
  \EndFor
\end{algorithmic}
\end{algorithm}
In Figure~\ref{fig:reduction}, a diagram of the Mask Reduction
Algorithm, Algorithm~\ref{alg:mreduce}, is presented.
The algorithm proceeds as follows.  For each patch in the phase space
hierarchy, the data on the patch is first refined in those directions that will
remain after the reduction.  The \new{covered} regions are then masked out;
this is accomplished by using masks that have been precomputed using a modified
communication algorithm\footnote{The mask variable is set to unity
  everywhere.  Then a standard communication algorithm
  from fine to coarse levels is executed on the mask variable in the
  phase space hierarchy, but instead of copying fine data to coarse
  cells where overlap occurs, zeros are copied into the coarse cells.} 
and stored in the phase space hierarchy.  A one-to-one
mapping is used to obtain the configuration-space partial-sum patch
$\mathcal{P}_{p}$ from
the pre-computed partial summation level $\mathcal{L}_{\text{partial}}$
that corresponds to the current 
phase space patch $\mathcal{P}$.  The summation is then executed, and
the results are 
placed in the 
configuration-space partial summation patch $\mathcal{P}_{p}$.  Once all
patches in the 
phase space hierarchy $\mathcal{G}_H$ have been reduced, a communication
from the partial summation level $\mathcal{L}_{\text{partial}}$ to the
total reduction level $\mathcal{L}_{\text{total}}$ is executed using
an accumulation operation.\footnote{Instead of copying values from each
  source patch to each destination patch, values are added from each
  source patch to each destination patch.}
The total reduction level at this point contains the total reduction on
a set of disjoint patches at the finest resolution.  A standard
communication operation from the total reduction level
$\mathcal{L}_{\text{total}}$ to the
configuration space hierarchy $\mathcal{G}^c_H$ completes the data transfer. 

We note that the total reduction level is not necessary.  One could
communicate directly between the partial summation level and the
configuration space hierarchy using an accumulation operation.  However,
for clarity and ease of implementation, we 
favored the use of an intermediate total reduction level.

While the Mask Reduction Algorithm is simple to implement, one might
suspect that it is not as efficient as it could be because interpolation
is done in phase space to the original data.  Instead, consider
execution of the reduction on the composite grid\new{,
  $\mathcal{G}_C$. Let $P_C$ be the number of patches in $\mathcal{G}_C$.
I}n a single dimension:
\begin{equation}
\begin{split}
  \rho_{\mbi_c} &=
 1-V_v\sum\limits_{d=N}^{N+M-1}\sum\limits_{j_d=0}^{J_d-1}f_{\mbi}
 = 1-V_v
 \sum\limits_{p=\new{0}}^{\new{P_C-1}}\sum\limits_{d=N}^{N+M-1}\sum\limits_{j_d=j^p_{d,\text{lo}}}^{j^p_{d,\text{hi}}}\prod\limits_{d'=0}^{N-1}\mathcal{I}_{j_{d'}}^{l(p),L\new{-1}}f^{p}_{\mbi},\\
& = 1-V_v
 \sum\limits_{p=\new{0}}^{\new{P_C-1}}\prod\limits_{d'=0}^{N-1}\mathcal{I}_{j_{d'}}^{l(p),L\new{-1}}\sum\limits_{d=N}^{N+M-1}\sum\limits_{j_d=j^p_{d,\text{lo}}}^{j^p_{d,\text{hi}}}f^{p}_{\mbi},\\
&= 1-
 \sum\limits_{p=\new{0}}^{\new{P_C-1}}\prod\limits_{d'=0}^{N-1}\mathcal{I}_{j_{d'}}^{l(p),L\new{-1}}
 \rho^P_{\mbi_c},
\end{split}
\end{equation}
where, since the composite grid has no levels, the level index $l$ is
a function of the patch index $p$ and is merely a label indicating the
refinement relative to the coarsest patches. 
Note the savings of the last step; instead of applying the potentially
costly prolongation operator at every grid level $j$, it is instead
applied to the lower-dimensional partial sums on each patch.  To achieve
this simplification, however, an efficient algorithm is needed to
construct the composite grid.

\begin{algorithm}[H]
\caption{Subdivision of patch into composite grid sub-patches}
\label{alg:subdiv}
\begin{algorithmic}
  \Procedure{ComputeSubPatches}{$\mathcal{P}_{\text{in}},\mathcal{L}_{\text{in}},\mathcal{G}_H$}
  \ForAll{levels $\mathcal{L}\in\mathcal{G}_H:\mathcal{L}>\mathcal{L}_{\text{in}}$}
  \State $\mathcal{S}\leftarrow \mathcal{P}_{\text{in}}$
  \ForAll{patches $\mathcal{P}\in \mathcal{L}:\mathcal{P}\cap\mathcal{P}_{\text{in}}\ne\emptyset$}
  \State $\mathcal{P}_{\text{extend}}\leftarrow\prod_{d=N}^{N+M-1}\repl_d\left((\mbi^{\text{in}}_{\text{lo}})_d,(\mbi^{\text{in}}_{\text{hi}})_d\right)\mathcal{P}$
  \ForAll{patches $\mathcal{P}_s\in\mathcal{S}$}
  \State $\mathcal{S}\leftarrow\mathcal{S} - \mathcal{P}_s + \mathcal{P}_s\cap \mathcal{P}_{\text{extend}} + \mathcal{P}_s\backslash(\mathcal{P}_s\cap \mathcal{P}_{\text{extend}})$  
  \State $\mathcal{S}\leftarrow\mathcal{S}-\mathcal{P}$
  \EndFor
  \EndFor
  \EndFor
  \State \Return $\mathcal{S}$
  \EndProcedure
\end{algorithmic}
\end{algorithm}
Such a procedure based on basic box calculus operations is presented in
Algorithm~\ref{alg:subdiv}.  Given a patch $\mathcal{P}_{\text{in}}$ and
the hierarchy in which it resides, a set of sub-patches $\mathcal{S}$ is
to be constructed.  Initially, the set of sub-patches is just the
original patch $\mathcal{P}_{\text{in}}$.  All patches
$\mathcal{P}$ from finer levels that overlap the patch are found.  In
the SAMRAI library, these relationships are already known and can be
obtained directly without searching.  Each overlapping patch, then, is
extended in the reduction directions, for example, that is, \new{its}
lower and upper indices in the \new{reduction} directions are replaced by
the lower and upper limits of the input patch $\mathcal{P}_{\text{in}}$:
\begin{equation}
\mathcal{P}_{\text{extend}}=\prod_{d=N}^{N+M-1}\repl_d\left((\mbi^{\text{in}}_{\text{lo}})_d,(\mbi^{\text{in}}_{\text{hi}})_d\right)\mathcal{P}.
\end{equation}
The extended overlapping patch $\mathcal{P}_{\text{extend}}$ is then
intersected with each patch $\mathcal{P}_s\in\mathcal{S}$, and both
the intersections and the complements replace the patch $\mathcal{P}_s$
in the set.  The original overlap patch $\mathcal{P}$  
is then subtracted from the set $\mathcal{S}$.  The extension ensures
that sub-patches of the greatest extent in the reduction directions can
be formed and that subsequent removal of overlap patches results in rectangular
sub-domains.  

\begin{figure}[t]
  \centering
  \includegraphics*[width=4.in]{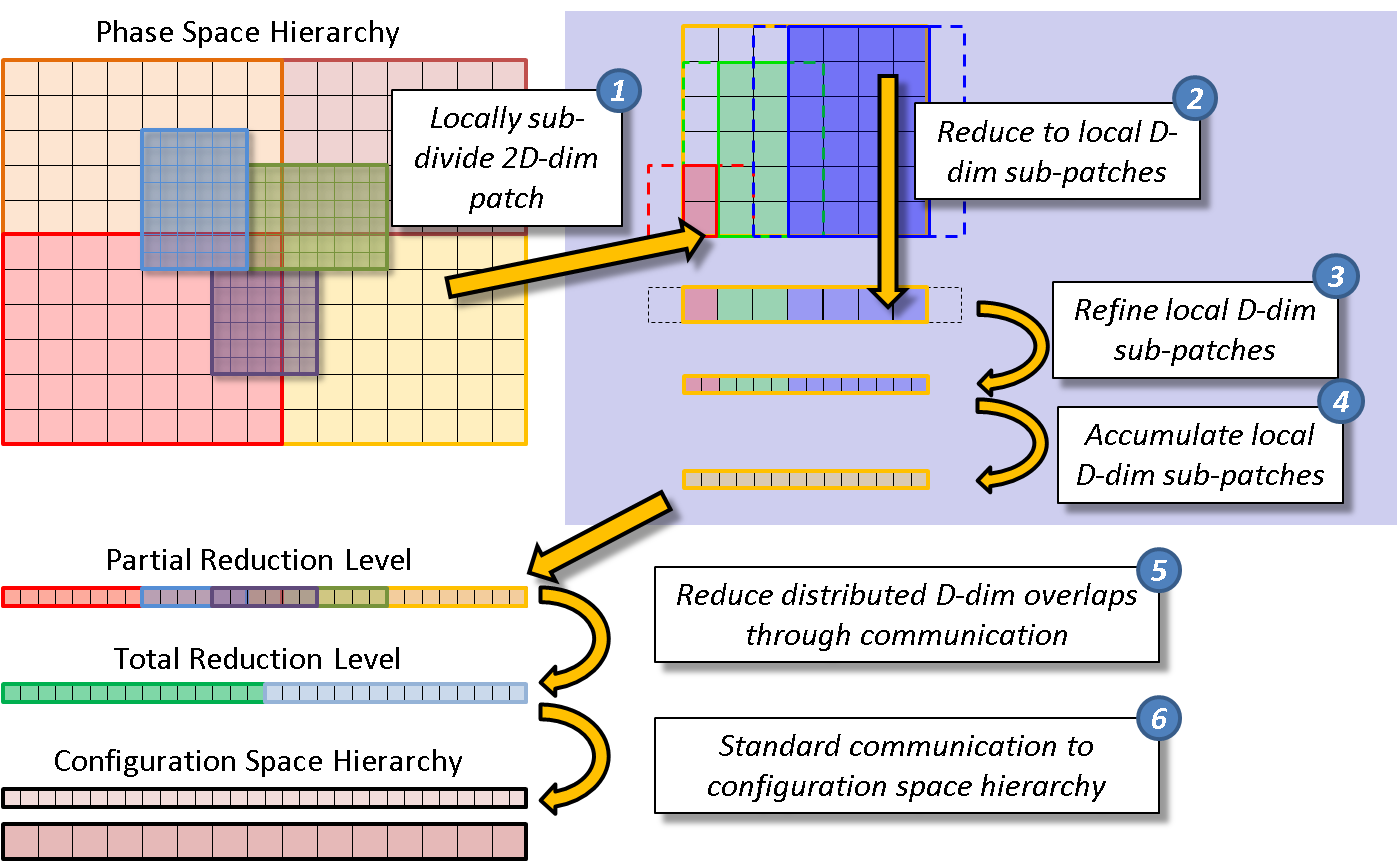}
  \caption{Graphical depiction of the Sub-Patch Reduction Algorithm.
  It is those steps in the light purple box that differ from the Mask
  Reduction Algorithm.}
  \label{fig:newreduction}
\end{figure}
\begin{algorithm}[H]
\caption{Sub-Patch Reduction}
\label{alg:spreduce}
\begin{algorithmic}
  \ForAll{levels $\mathcal{L}\in\mathcal{G}_H$}
  \ForAll{patches $\mathcal{P}\in \mathcal{L}$}
  \State Find patch $\mathcal{P}_{\text{p}}(\mathcal{P})\in\mathcal{L}_{\text{partial}}$ corresponding to patch $\mathcal{P}$
  \State $\mathcal{S}\leftarrow$ \textsc{ComputeSubPatches}($\mathcal{P},\mathcal{L},\mathcal{G}_H$)
  \ForAll{sub-patches $\mathcal{P}_s\in\mathcal{S}$}
  \State Grow ghost cells in non-reduction directions 
  \State Sum reduce data, including ghost cells, to temporary patch $\mathcal{P}_{\text{tmp}}$
  \State Refine data on $\mathcal{P}_{\text{tmp}}$ in non-reduction directions
  \State Copy data on interior of $\mathcal{P}_{\text{tmp}}$ to partial sum patch $\mathcal{P}_{\text{p}}$
  \EndFor
  \EndFor
  \EndFor
  \ForAll{patches $\mathcal{P}_{\text{t}}\in \mathcal{L}_{\text{total}}$}
  \State Accumulate data from co-located patches $\mathcal{P}_{\text{p}}$
  \EndFor
  \ForAll{levels $\mathcal{L}\in\mathcal{G}^c_H$}
  \ForAll{patches $\mathcal{P}\in \mathcal{L}$}
  \State Copy from co-located patches $\mathcal{P}_{\text{t}}$
  \EndFor
  \EndFor
\end{algorithmic}
\end{algorithm}
Using this sub-patch construction procedure, the more efficient Sub-Patch
Reduction Algorithm presented in Algorithm~\ref{alg:spreduce} and
depicted in Figure~\ref{fig:newreduction} can be
used.  As before, loops are performed over all patches, but now, for a
given patch, the set of sub-patches are identified.  Each of these
sub-patches is first grown in the non-reduction directions by the number
of ghost cells necessary for any subsequent prolongation operations.
The data on each sub-patch, including the ghost cells, is sum reduced to
a temporary configuration-space patch of the same resolution.  This
partial sum data is then refined, and the result is copied into the
corresponding partial sum patch from the partial sum hierarchy.  Once
all contributions from all sub-patches are obtained, the reduction
algorithm proceeds as before.

Finally, we note that neither reduction algorithm assumes that either
the phase-space or configuration-space hierarchies have a special
structure.  Because intermediate data structures are used along with 
standard communication algorithms, arbitrary meshes could be used in
either dimensions.  In addition, these algorithms are applicable in
parallel and for arbitrary dimension.

\subsubsection{Injection algorithm}
The process of injection \new{from} lower to higher dimensions is much simpler.
The data, $E_i$, for instance, is the same for all phase space locations
at index $i$, that is, $\hat{E}_{ij}=E_i$.  Nevertheless, to facilitate
the data transfer between hierarchies of different dimensions, it is
convenient to first construct an intermediate configuration-space
\emph{restricted hierarchy}, $\mathcal{G}^r_H=\{\mathcal{L}^r_l,\new{0}\le
l\le\new{L-1}\}$, where
\begin{equation}
  \mathcal{L}^r=\left\{\prod_{d=N}^{N+M-1}\restr_d\mathcal{P},\quad\forall\mathcal{P}\in\mathcal{G}_H\right\},
\end{equation}
that is, it is composed of lower-dimensional restrictions of all of the
patches in the phase space hierarchy $\mathcal{G}_H$.   

\begin{figure}[t]
  \centering
  \includegraphics*[width=4.in]{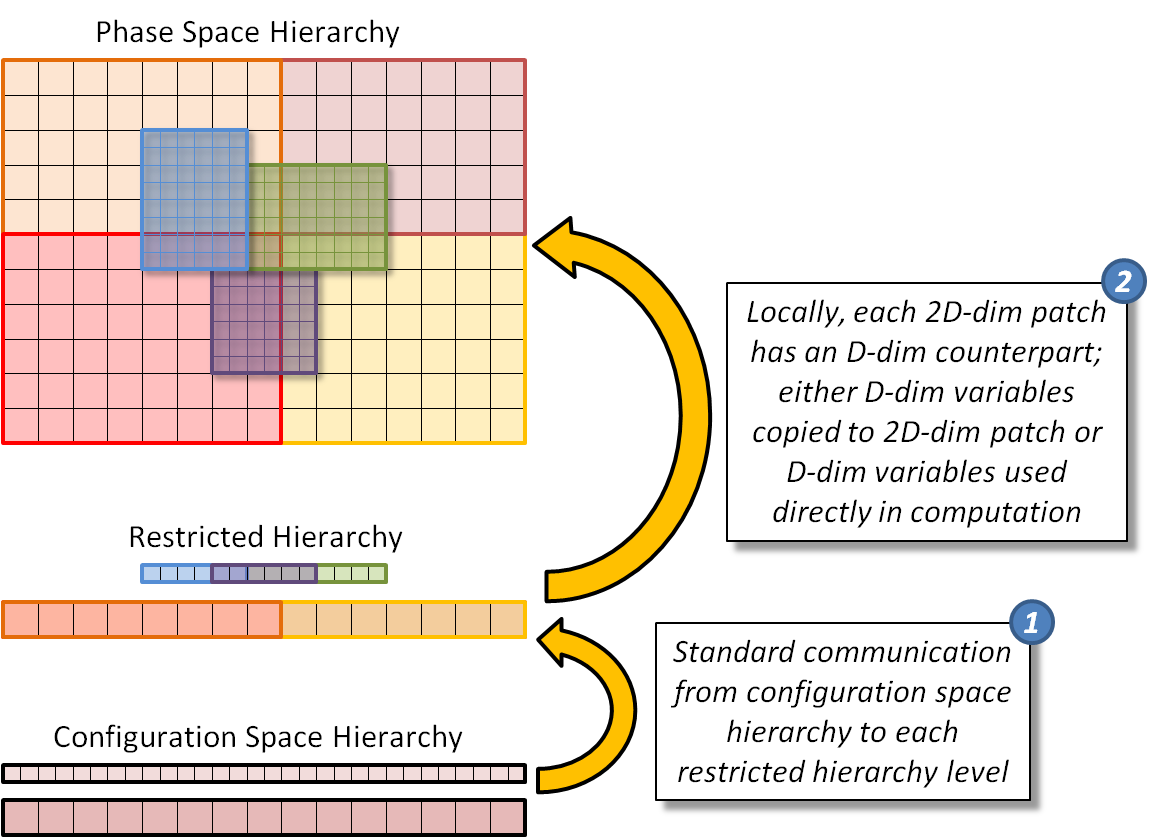}
  \caption{Injection}
  \label{fig:injection}
\end{figure}
\begin{algorithm}[H]
\caption{Injection}
\label{alg:inject}
\begin{algorithmic}
  \ForAll{Levels $\mathcal{L}\in\mathcal{G}^r_H$}
  \ForAll{Patches $p\in \mathcal{L}$}
  \State copy from co-located patches $p_c\in\mathcal{G}^c_H$
  \EndFor
  \EndFor
  \ForAll{Levels $\mathcal{L}\in\mathcal{G}_H$}
  \ForAll{Patches $p\in \mathcal{L}$}
  \State find patch $p_{\text{r}}(p)\in\mathcal{G}^r_H$ corresponding to patch $p$
  \State either copy from $p_{\text{r}}$ into $p$ or use directly
  \EndFor
  \EndFor
\end{algorithmic}
\end{algorithm}
For completeness, the injection transfer algorithm is depicted in
Figure~\ref{fig:injection} and presented in Algorithm~\ref{alg:inject}.
Standard communication copiers are used to fill the restricted hierarchy
$\mathcal{G}^r_H$ from the configuration space hierarchy
$\mathcal{G}^c_H$.  A one-to-one mapping exists from every restricted
hierarchy patch to the corresponding phase space hierarchy patch.  The
restricted hierarchy data can then be injected into phase space data,
\eg, $\hat{E}_{ij}\leftarrow E_i$.  This wastes storage with many
repeated values, so in the \Valhalla code, we directly access the
restricted hierarchy data when needed.

\subsection{Regridding \new{for multiple AMR hierarcies}}
\label{sec:regrid}
Working with multiple hierarchies introduces regridding
challenges for AMR algorithms.  With a single hierarchy, the user
typically defines a regrid frequency.  At user-defined times,
refinement criteria are used to identify cells in need of refinement
(coarsening), new levels of rectangular patches are formed containing
these flagged cells and are populated, and these new levels replace old
levels in the hierarchy.  With multiple hierarchies, one must decide to
what degree to constrain the regridding of each hierarchy.
Considerations include the facilitation of efficient communication
between hierarchies, the cost/benefit of adaptation for each hierarchy,
and the degree \new{and nature of dependencies between hierarchies
  (\eg, can the hierarchies refine independently, and if not, are
    the dependencies one-way or more complicated?)}  In
the case of Vlasov simulation, \new{coordination} is most critical
between the configuration space hierarchy and the phase 
space hierarchies, where a variety of intermediate data structures are
required to execute inter-dimensional data transfer.

For the purposes of demonstrating proof-of-principle, we made
several simplifying choices.  For 1D+1V Vlasov-Poisson, the
electrostatic problem is solved in 1D.  When restricted down to 1D, mesh
refinement of features in 2D, such as particle trapping regions, 
will typically lead to mesh refinement almost everywhere; 
hence, there is little advantage to mesh refinement in configuration
space.  Furthermore, the cost of the higher dimensional solve by far
dominates the total cost, so there should be little advantage to
using mesh refinement to speed-up the lower-dimensional solve.  We
therefore elected to require the configuration space mesh to 
be uniform at the finest level of refinement of the phase space
hierarchy.  While the mesh was decomposed into patches, we did not
distribute the configuration space hierarchy across multiple
processors.  \new{For higher-dimensional
  problems, such as 2D+2V Vlasov-Poisson, one may benefit from distributing the
  configuration-space hierarchy. However, such a distribution will be over
  a much smaller number of processors than the phase-space hierarchy
  simply because there is so much less data and work to distribute in
  lower dimensions.\footnote{\new{For a uniform-grid 2D+2V Vlasov-Poisson
    code, we have seen in practice that the Poisson solve benefits from
    distributed parallelism only when the problem size has grown such
    that the Vlasov solve occurs across several thousand processors.}}} 
When the phase-space hierarchy was regridded, the 
configuration-space hierarchy was only regridded if a new level of refinement was
added (removed) in phase space.  This scheme had a secondary advantage of
simplifying the Poisson solver; solves were executed on the uniform,
finest mesh level in configuration space and then averaged down to
coarser levels, thereby avoiding the need for a Fast Adaptive Composite
(FAC) iterative algorithm~\cite{McTo86}.  We note that these are merely choices
and not requirements; the algorithms for
inter-hierarchy data transfers defined in Section~\ref{sec:moment} 
support more general configuration and phase-space hierarchies.

\subsubsection{Managing inter-hierarchy \new{coordination}}
\begin{figure}[t]
  \centering
  \includegraphics*[width=3.5in]{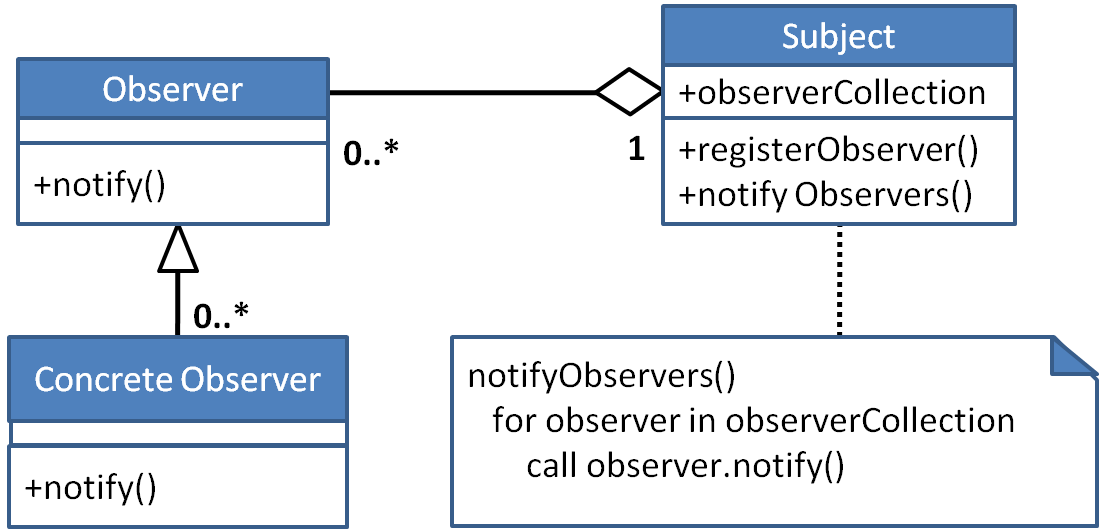}
  \caption{Unified Modeling Language depiction of the Observer Design
  Pattern used to create notifying hierarchies.  In our case, the
  reduction and injection algorithms for configuration-space are
   \texttt{HierarchyObserver}s.  These observers register themselves with the 
  phase space  \texttt{NotifyingHierarchy} to receive notices about regridding.}
  \label{fig:nothierarchy}
\end{figure}
From the descriptions of the reduction and injection transfer
algorithms in Section~\ref{sec:moment}, it is clear that the
intermediate data structures, such as the partial sum level or
restricted patch hierarchy, are  
dependent on the phase and configuration space hierarchies.  When regridding
of any of the primary hierarchies occurs, the intermediate data
structures must be rebuilt in order to maintain consistency.  To
facilitate this, we made use of the Observer Design
Pattern~\cite{Holub2004} as depicted in Figure~\ref{fig:nothierarchy}.  The
SAMRAI concept of \texttt{PatchHierarchy} was generalized to allow other
objects,  such as the \texttt{ReductionAlgorithm} and
\texttt{InjectionAlgorithm}, to subscribe 
to the phase space hierarchy in order to receive messages indicating
that the phase space hierarchy had regridded.  Reconstruction of the 
intermediate data structures is deferred until a subsequent reduction or
injection operation is attempted and a message from an observed
hierarchy is found.

\subsubsection{Mesh Refinement Criteria}
Finally, selection of mesh refinement criteria can be critical in
obtaining optimum AMR performance.  For our purposes here, we 
chose to apply common heuristic refinement criteria to the phase space
distribution function.  Specifically, we tag cells when
\begin{equation}
  \delta_1f_{\mbi}+\delta_2f_{\mbi}>\text{tol}
\end{equation}
where
\begin{equation}
  \delta_1f_{\mbi}=\left[\half\sum_{d=1}^D\Delta{x}_d(f_{\mbi+\mbe^d}-f_{\mbi-\mbe^d})^2\right]^{\half}
  \quad\text{and}\quad 
  \delta_2f_{\mbi}=\half\sum_{d=1}^D\Delta{x}_d^2|f_{\mbi+\mbe^d}-2f_{\mbi}+f_{\mbi-\mbe^d}|
\end{equation}
estimate the first two truncation error terms.
We do not claim that this is the optimal choice; it is merely sufficient
to demonstrate our algorithms.  Other error indicators could be used,
including indicators based on physical principles, such as local
estimates of the location of the trapped-passing boundary.  The choice
of optimal refinement criteria is intimately related to  problem-specific
quantities of interest, so we leave this topic for future work.

\section{Numerical Results}
\label{sec:results}
We present results from a \Valhalla simulation
of the bump-on-tail instability~\cite[\S 9.4]{KrTr73} \new{as a
basic proof-of-principle of the block-structured AMR approach for
Vlasov-Poisson simulation}.  We used the same
problem specified in our previous discretization work~\cite{BaHi10}.
The initial distribution function was given by  
\begin{equation}
  f = f_{b}(v) \left(1+0.04\cos\left(0.3x\right)\right),
\end{equation}
with
\begin{equation}
  f_{b}(v) = \frac{0.9}{\sqrt{2\pi}}\exp{\left(-\frac{v^2}{2}\right)}+\frac{0.2}{\sqrt{2\pi}}\exp{\left(-{4(v-4.5)^2}\right)}.
\end{equation}
The $(x,v)$ domain was $[-10\pi/3,10\pi/3]\times[-8,10]$ and was periodic
in the $x$-direction.  We initialized the solution with a coarse grid of
$N_x\times N_v=16\times32$ and with an initial refinement in the
box $[(0,8),(15,24)]$.  This initial mesh configuration allowed for
larger time steps, since the cells along the maximum velocity boundary
have a larger aspect ratio.  \new{The initial time step was $\Delta
t_0=0.01$, and this was allowed to adjust to 50\% of the local
stability condition based on the linear stability of the fourth-order scheme
(See~\cite{CoDoHiMa10}).  Time steps could increase no more than 10\%
from their previous value, but could decrease by any amount.}
The grid refinement
criteria tolerance was $\text{tol}=0.01$, and grid refinement ratios of
$\mbr_0^1=[2,4]$, $\mbr_1^2=[4,2]$, and $\mbr_2^3=[2,2]$
were used.  \new{Up to four levels of AMR mesh were allowed.  To isolate
  the AMR performance issues, we consider the 
  serial performance on a single node of the LLNL 64-bit AMD Linux
  cluster \texttt{hera}.

AMR performance is very problem-dependent.}  When small regions of
refinement are required, in particular, when there are lower-dimensional
features in the solution, 
AMR is generally a net win.  However, there is overhead associated with
AMR for which sufficient problem size reduction is necessary to achieve
a net gain in simulation performance.  Performance is also highly dependent
on the choice of parameters, such as regrid frequency and refinement
tolerances, so the results presented here are meant to demonstrate that
our Vlasov-AMR procedure works and can show savings.  Whether or not AMR is
useful in other specific cases and optimal choices for AMR parameters
and regridding criteria are very important issues.

\begin{table}
\centering
\new{
\begin{tabular}{|c||c|c|c|}\hline
\textbf{Parameter} & \textbf{AMR1} & \textbf{AMR2} & \textbf{AMR3}\\\hline\hline
\texttt{largest\_patch\_size} & & & \\
\hspace{2em}\texttt{level\_0} & (32,32) & (16,32) & (16,32) \\
\hspace{2em}\texttt{level\_1} & (64,64) & (32,128) & (32,128) \\
\hspace{2em}\texttt{level\_2} & (64,64) & (128,256) & (128,256) \\\hline
\texttt{smallest\_patch\_size} & (4,4) & (8,8) & (8,8) \\\hline
\texttt{regrid\_interval} & 2 & 4 & 8 \\\hline
\texttt{tag\_buffer} & (1,1) & (4,4) & (8,8)\\\hline
\end{tabular}
}
\label{tab:amr_param}
\caption{\new{AMR parameters used to define the three test configurations.
  The parameters \texttt{largest\_patch\_size} and
  \texttt{smallest\_patch\_size} control the largest and smallest allowable
  patch sizes level-by-level; if unspecified, the finest specified value
  is applied to all subsequent levels.  The parameter
  \texttt{regrid\_interval} is the frequency, in time steps, at which
  regridding occurs.  Finally, \texttt{tag\_buffer} is the number of
  buffers cells to add around a 
  region tagged for refinement to facilitate less frequent regridding.}}
\end{table}
\new{To help elucidate the AMR performance, we considered three AMR
  parameter configurations, as shown in Table~\ref{tab:amr_param}.  The
  AMR1 case represents an attempt to minimize the number of refined
  cells by using smaller patches and more frequent regridding.  The AMR2
  and AMR3 cases allow for larger patches an less frequent regridding.
  Thus, these three cases can give some sense of the trade-offs between
  reducing the amount of mesh (AMR memory reduction) and reducing the
  run time (AMR speed-up).  All cases were run using the \emph{Sub-Patch
  Reduction} algorithm with unlimited fifth-order reconstruction unless
  otherwise noted.}

\begin{figure}[t]
  \centering
  \hfil\includegraphics*[width=2.99in]{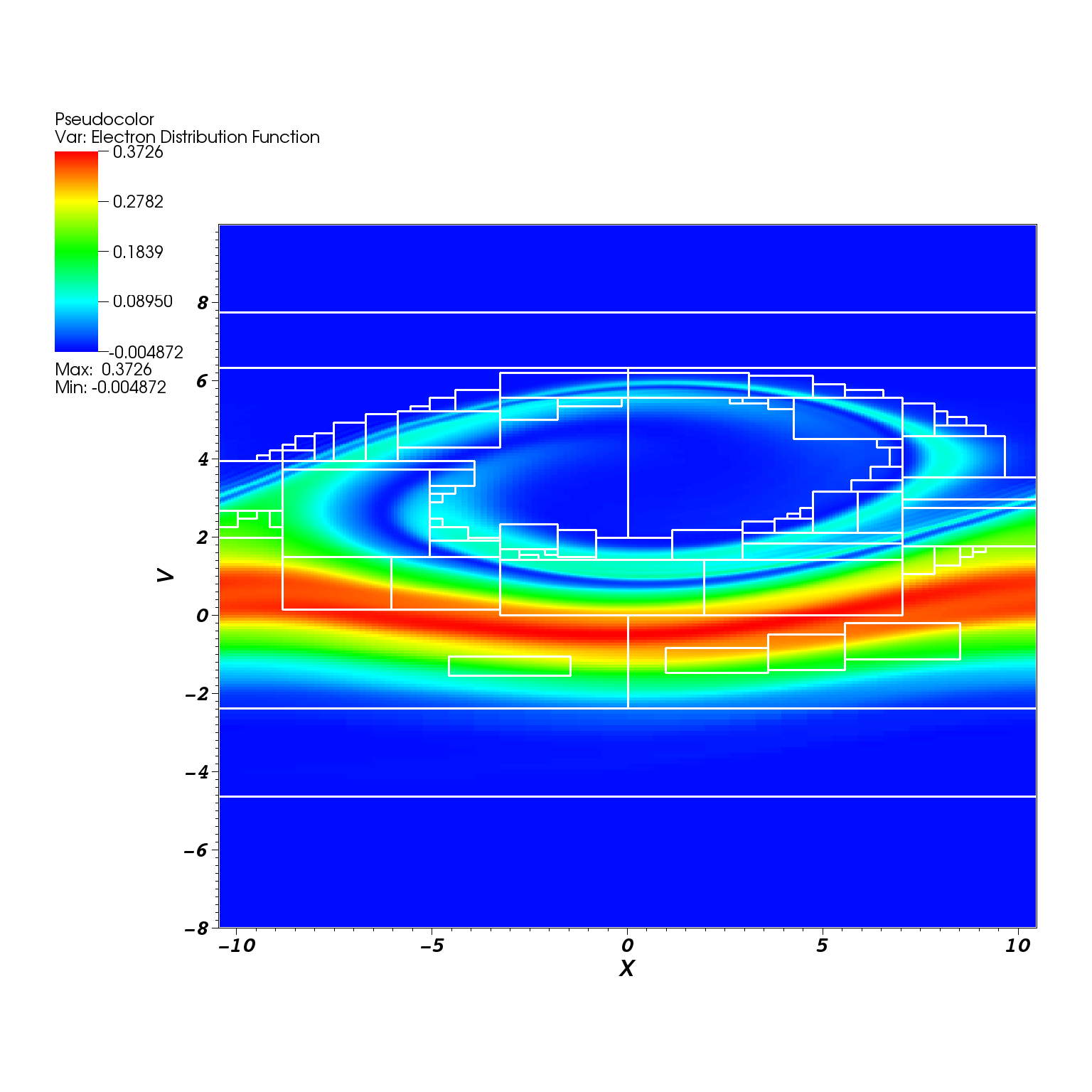}\hfil
  \includegraphics*[width=2.99in]{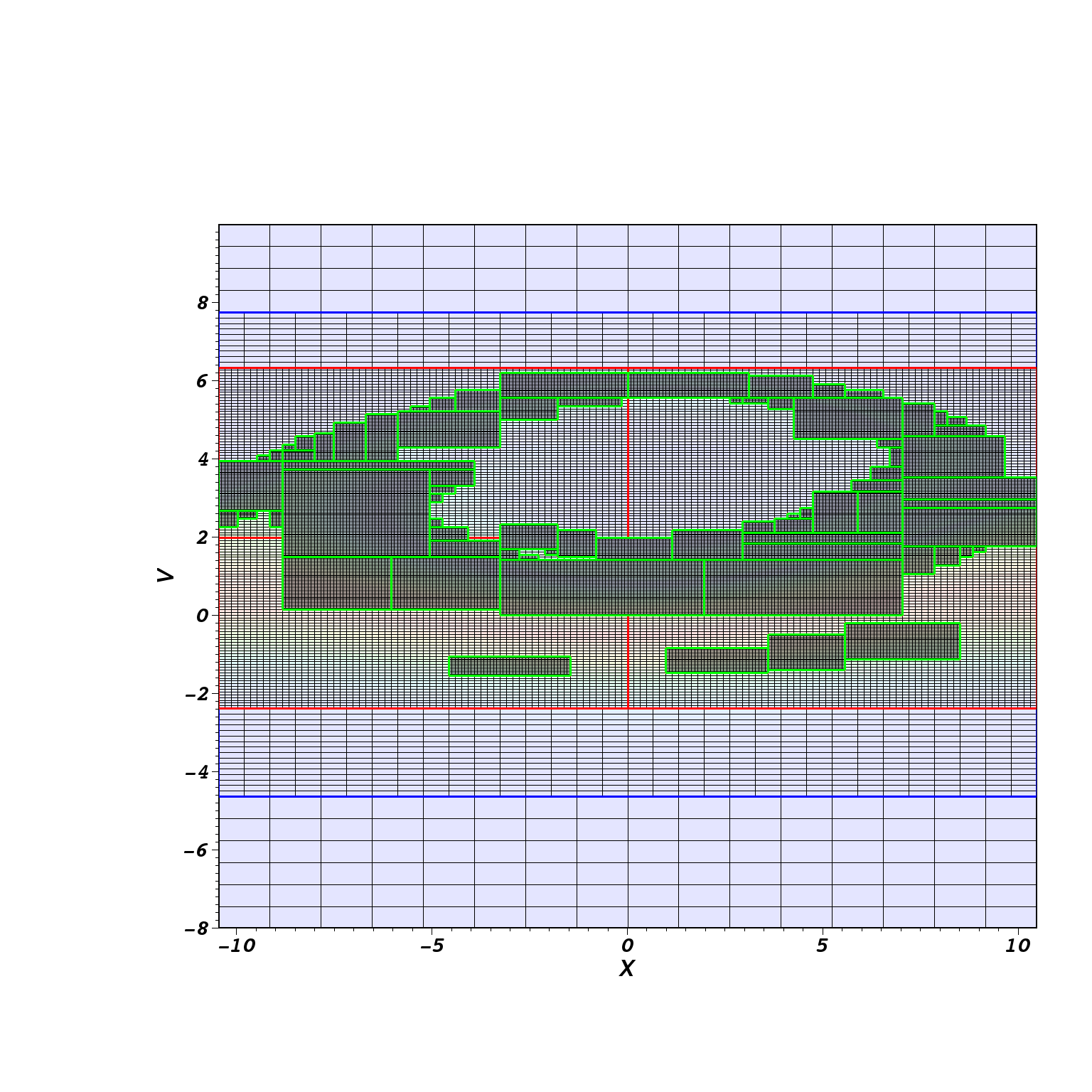}\hfil
  \caption{\new{Example result} for the bump-on-tail problem at time
  $t=22.5$ \new{for the AMR1 case}. \new{On the left, the distribution
      function is shown.  Boxes 
      outlined in white denote AMR patches.  On the right,} the corresponding
    four-level AMR mesh \new{is} shown.  The mesh adapts to resolve the
    particle trapping region as it forms.  \new{Note that the minimum
      distribution function is small but negative; no positivity
      enforcement schemes were used in this calculation.}} 
  \label{fig:amrsol}
\end{figure}
\begin{figure}[t]
  \centering
  \includegraphics*[width=4.in]{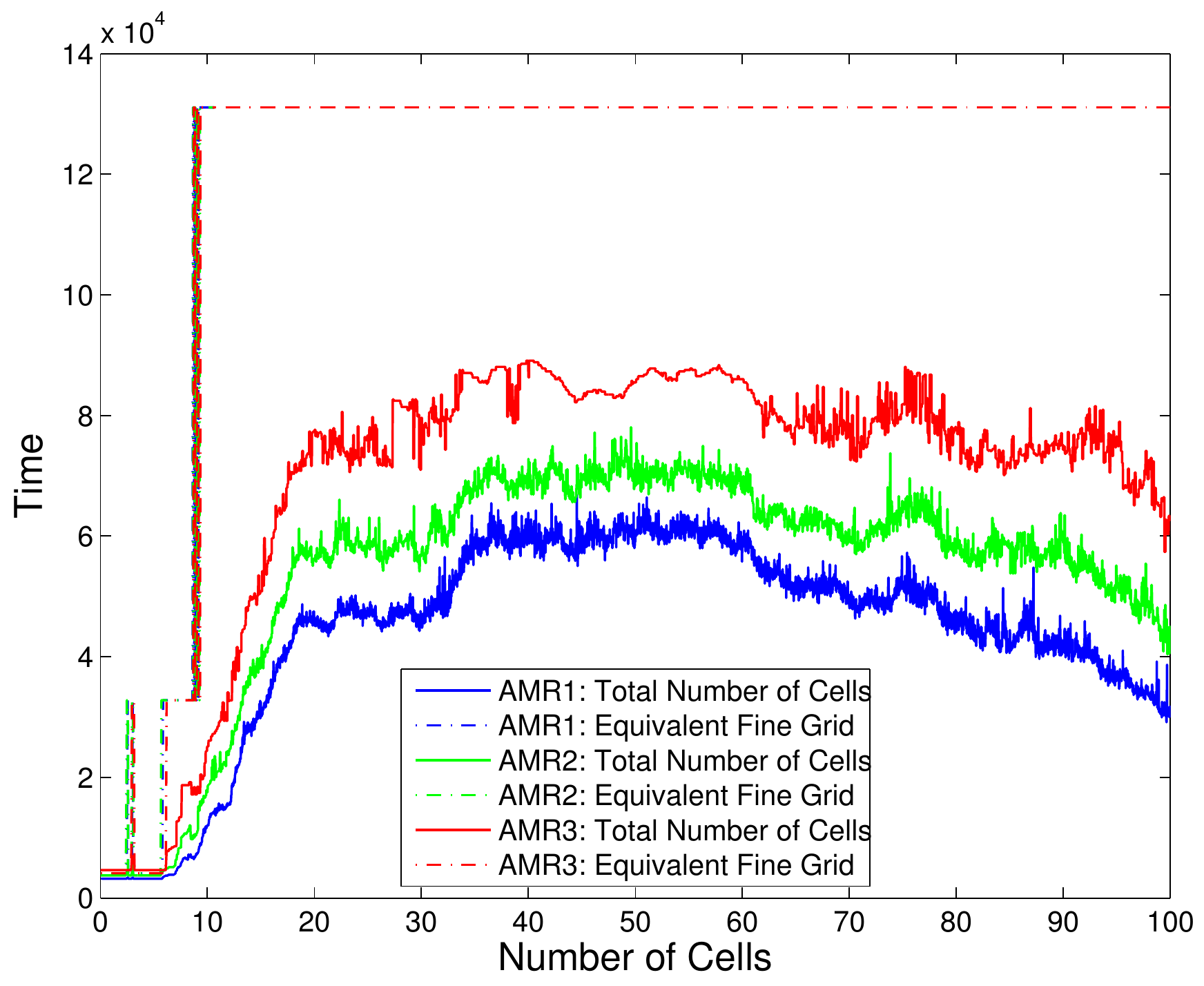}
  \caption{Time history of the number of cells for the bump-on-tail
  simulation \new{for the three AMR parameter configurations}.  The
  \new{dashed} curve\new{s are} the number of cells in an 
  equivalent uniform grid based on the current maximum refinement level,
  while the \new{solid} curve\new{s are} the actual number of cells in
  the AMR hierarchy.  \new{Note that around $t=3, 6$ and $9$, there is
    some intermittency in all cases, as the adaptivity adds and removes
    a small number of patches at the next finer level.}} 
  \label{fig:ncells}
\end{figure}
Figure~\ref{fig:amrsol} shows computed approximations of the
phase-space distribution function at $t=22.5$ \new{for case AMR1}.  As
expected, we see a concentration of the mesh only in regions of most
rapid variation in the solution, and conversely, we see mesh coarsening
in the the trapping region.
At this point in the calculation, the total number of
cells is 40784, compared to 131072 cells in the
equivalent fine grid -- a reduction of approximately 69\%. 
In Figure~\ref{fig:ncells}, we show the
time history of the number of AMR cells plotted against the
instantaneous equivalent fine grid {for each AMR parameter
  configuration}.  We see that once adaptivity starts, we can achieve
a\new{n average} reduction of \new{between forty and sixty percent,
  depending on the AMR parameters.  As expected, the AMR1 case uses the least
  number of cells, and the AMR3 case, because of its increased patch
  size and tagging buffer, uses the most cells.  Considering
  Figure~\ref{fig:amrsol}, we see that a lot of the mesh reduction comes
  in the velocity (vertical) dimension, and this is expected for each
  additional velocity dimension in higher dimensions.  In 1D+1V,
  there is little localization in configuration space.  However, in
  2D+2V, there is also the opportunity for spatial localization of the
  features, which would result in even more mesh reduction.}

\begin{figure}[t]
  \centering
  \includegraphics*[height=2.4in]{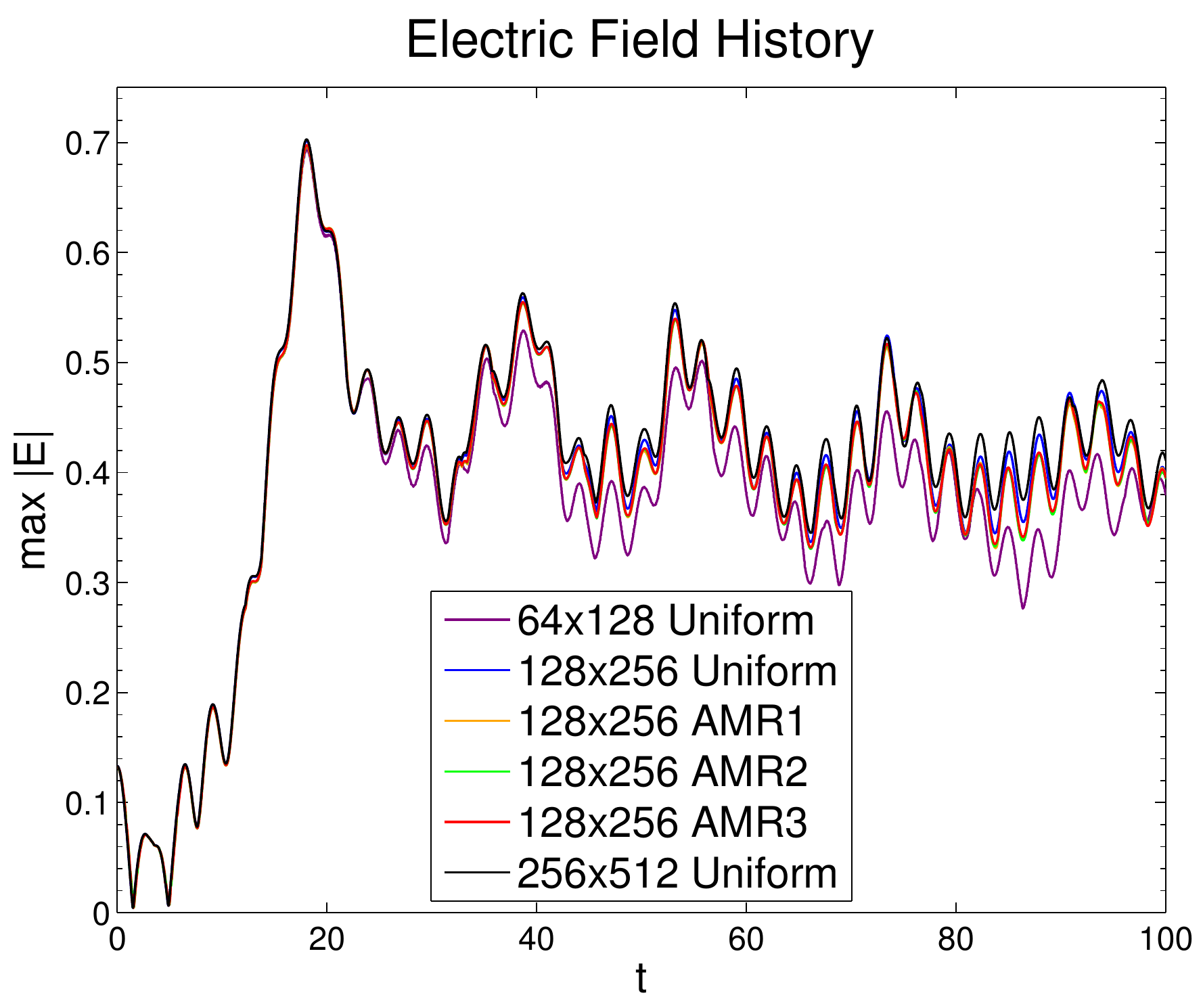}\hfil
  \includegraphics*[height=2.4in]{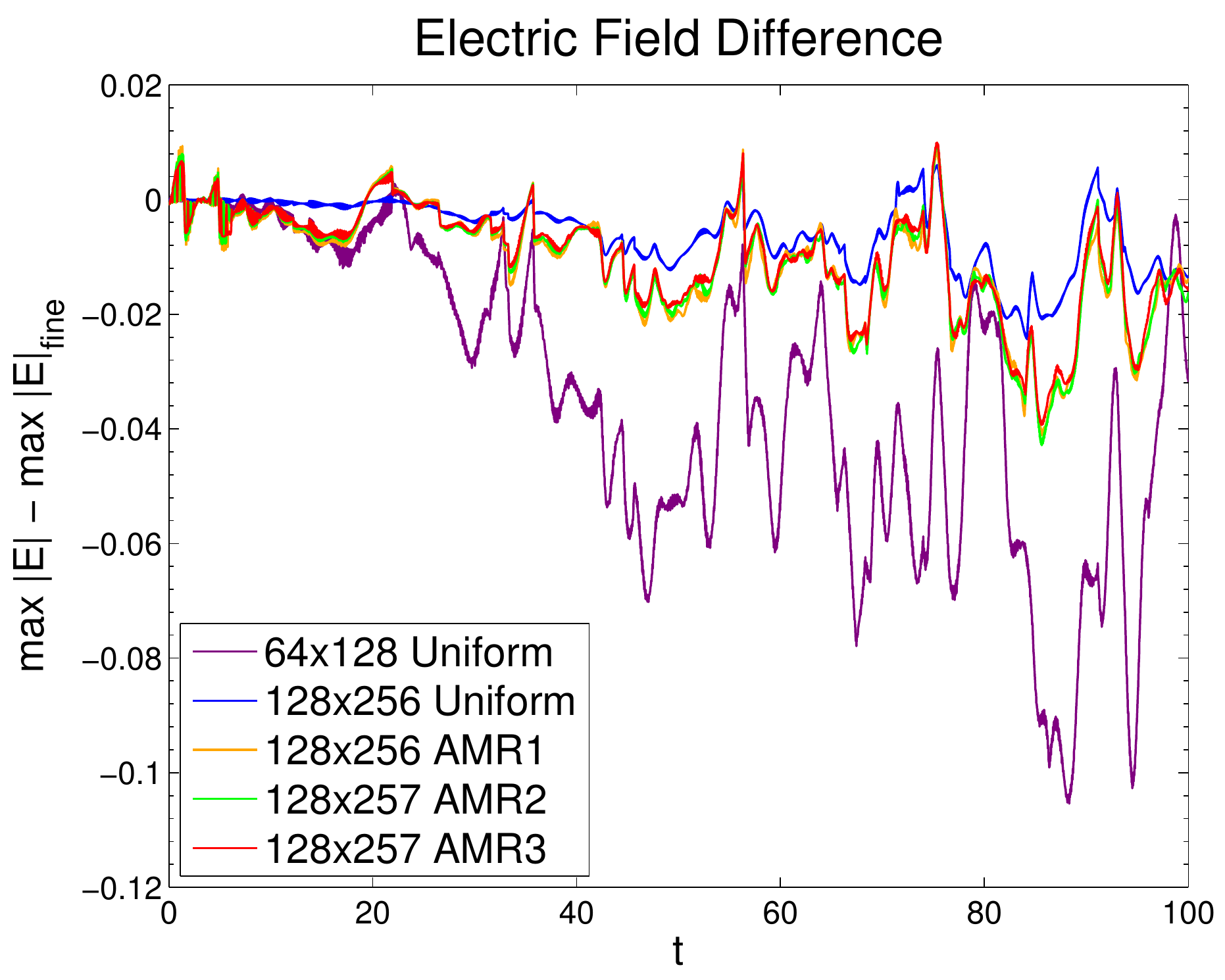}
  \caption{Time history of the maximum electric field for bump-on-tail
  calculations \new{at several resolutions}.  \new{Results are from the
    three AMR configurations in which the finest resolution is
    equivalent to a $128\times256$ uniform
    mesh as well as reference uniform-grid calculations on meshes of
    $64\times128$, $128\times256$, and $256\times512$.  The left plot
    shows the maximum of the electric field.  The right plot is the same
    data, plotted as the difference from the results from the fine
    $256\times512$ mesh.}}
  \label{fig:efhistory}
\end{figure}
In Figure~\ref{fig:efhistory}, we present the time history of the
maximum of the electric field for the bump-on-tail problem for three
different resolutions.  \new{This metric is a fairly sensitive measure
  of numerical fidelity.  In addition to the three AMR parameter cases, we
  plot the results from three uniform-grid cases: $64\times\new{128}$,
  $128\times\new{256}$, and $256\times\new{512}$.  We use the finest of
  these as a ``reference'' solution to plot the discrepancy of the
  electric field maximum.  For the AMR calculations, the finest local
  resolution is equivalent to the $128\times256$ uniform mesh.  At
  early times, when the solution has little structure, all of the
  solutions agree well.  The small up/down differences in the AMR results
  before $t=10$ are due to the discrete temporal resolution (the AMR cases use
  larger time steps) of the first two valleys of the maximum electric field.
  Around $t=25$, one begins to see significant differences in the
  coarsest solution, since it cannot resolve as well the features being
  generated in the particle trapping region. We can conclude from these 
  results that the $64\times128$ resolution was insufficient to
  accurately track the maximum electric field over this time interval;
  thus the increased resolution of the AMR is necessary.

  By about $t=50$, one sees a growing discrepancy between all of the AMR
  cases and the equivalent uniform mesh of $128\times256$; over the
  interval considered, the discrepancy is roughly twice as large at its
  maximum.  One explanation for this could be the accumulation of error
  over longer integration times.  Another likely explanation is that we
  are not capturing all of the relevant details with the refined mesh because
  we are using a simple heuristic gradient detection algorithm; more 
  problem-specific refinement criteria may perform better.
  Nevertheless, the AMR results do track the equivalent uniform mesh
  results well.  Compared to the finest uniform grid results, the phase
  of the AMR results is relatively good, but the amplitude is being
  under-predicted by an increasing amount over time; there will, of
  course, be slightly more dissipation in the coarser results when
  features appear that cannot be adequately resolved.}
These results show that AMR can provide effectively equivalent results
as a uniform mesh.  \new{Of course, one must consider the quantities of
interest for the calculation, and suitable choices of AMR parameter and
refinement criteria need to be selected.}

\begin{table}
\centering
\new{
\begin{tabular}{|c||c|c|c|c|c|}\hline
\textbf{Case} & \textbf{Time 1 (s)} & \textbf{Time 2 (s)} & \textbf{Time
  3 (s)} & \textbf{Avg Time (s)} & \textbf{Speed-up} \\\hline\hline
Uniform & 1521 & 1520 & 1519 & 1520 & 1.00\\\hline
AMR1 & 2622 & 2615 & 2632 & 2623 & 0.58\\\hline
AMR2 & 1355 & 1346 & 1336 & 1346 & 1.13 \\\hline
AMR3 & 960 & 961 & 960 & 960 & 1.58\\\hline
\end{tabular}
}
\label{tab:speedup}
\caption{\new{AMR speed-up on the \texttt{hera} cluster for the three AMR
    parameter cases.  The average of three results for each case are
    compared to the average run time for three instances of the same
    problem solved using the equivalent uniform mesh.}}  
\end{table}
\new{
In addition to mesh reduction, the potential for decreased
time-to-solution by using AMR is also of interest.    As
indicated earlier, AMR should have the benefit that the equations are
integrated on fewer cells and that larger time steps can be taken.
However, traditional AMR incurs additional overhead from the
regridding and the data transfers (communication, interpolation, and
averaging) between patches on different levels.  The Vlasov-Poisson
system has additional overhead due to the reduction and injection
operations between different dimensions. 

In Table~\ref{tab:speedup}, we provide run times on the
\texttt{hera} cluster for the three AMR parameter
cases in comparison to the run time for the equivalent uniform mesh.
The AMR1 case, with refinement every other step, causes significant
slow-down of the code; however, in the other two cases, the time to
solution is reduced.  As expected, when the regridding frequency is
reduced, the speed-up increases.  We note that we have erred 
conservatively in favor of the uniform mesh solution; running serially
and with a single patch, it incurs no inter-patch communication costs.}

\begin{table}
\centering
\new{
\begin{tabular}%
{|l%
 |>{\PBS\raggedleft\hspace{0pt}}p{.8in}%
  >{\PBS\raggedright\hspace{0pt}}p{.8in}%
 |>{\PBS\raggedleft\hspace{0pt}}p{.8in}%
  >{\PBS\raggedright\hspace{0pt}}p{.8in}|}\hline
\textbf{Case} & \multicolumn{2}{|c|}{\textbf{Exclusive Time (s)}} & \multicolumn{2}{|c|}{\textbf{Total Time
  (s)}}\\\hline\hline
\multicolumn{5}{|l|}{\tt MultiStageIntegrator::computeRHS()}\\\hline
\texttt{Uniform} & 830.6 & (55\%) & 830.7 & (55\%) \\\hline
\texttt{AMR1} & 169.0 & (6.4\%) & 169.1 & (6.4\%) \\\hline
\texttt{AMR2} & 170.5 & (13\%) & 170.6 & (13\%) \\\hline
\texttt{AMR3} & 236.0 & (25\%) & 236.0 & (25\%) \\\hline\hline
\multicolumn{5}{|l|}{\tt MultiStageIntegrator::evaluateInstantaneousConstraints()}\\\hline
\texttt{Uniform} & 16.40 & (1.1\%) & 115.0 & (7.6\%) \\\hline
\texttt{AMR1} & 68.89 & (2.6\%) & 272.6 & (10\%) \\\hline
\texttt{AMR2} & 19.63 & (1.5\%) & 132.5 & (9.9\%) \\\hline
\texttt{AMR3} & 12.35 & (1.3\%) & 89.76 & (9.4\%) \\\hline\hline
\multicolumn{5}{|l|}{\tt xfer::RefineSchedule::fillData()}\\\hline
\texttt{Uniform} & 32.25 & (2.1\%) & 94.65 & (6.2\%) \\\hline
\texttt{AMR1} & 21.95 & (0.83\%) & 1564. & (59\%) \\\hline
\texttt{AMR2} & 14.54 & (1.1\%) & 751.9 & (56\%) \\\hline
\texttt{AMR3} & 14.00 & (1.5\%) & 384.3 & (40\%) \\\hline\hline
\multicolumn{5}{|l|}{\tt ReductionAlgorithm::reduce()}\\\hline
\texttt{Uniform} & 0.8683 & (0.06\%) & 76.56 & (5.0\%) \\\hline
\texttt{AMR1} & 7.507 & (0.29\%) & 153.5 & (5.8\%) \\\hline
\texttt{AMR2} & 3.179 & (0.24\%) & 74.75 & (5.6\%) \\\hline
\texttt{AMR3} & 1.846 & (0.19\%) & 43.31 & (4.5\%) \\\hline\hline
\multicolumn{5}{|l|}{\tt MultiStageIntegrator::regridHierarchies()}\\\hline
\texttt{Uniform} & 0.000 & (0\%) & 0.000 & (0\%) \\\hline
\texttt{AMR1} & 72.70 & (2.8\%) & 767.9 & (29\%) \\\hline
\texttt{AMR2} & 17.48 & (1.3\%) & 203.6 & (15\%) \\\hline
\texttt{AMR3} & 6.150 & (0.64\%) & 74.41 & (7.8\%) \\\hline\hline
\multicolumn{5}{|l|}{\tt ConservativeWENORefine::WENO\_2D}\\\hline
\texttt{Uniform} & 0.000 & (0\%) & 0.000 & (0\%) \\\hline
\texttt{AMR1} & 639.0 & (24\%) & 720.2 & (27\%) \\\hline
\texttt{AMR2} & 382.2 & (29\%) & 411.1 & (31\%) \\\hline
\texttt{AMR3} & 203.7 & (21\%) & 215.1 & (22\%) \\\hline
\end{tabular}
}
\label{tab:costs}
\caption{\new{A summary of the costs of six key routines for the three
    AMR cases and the equivalent uniform mesh.  Exclusive time is the
    time strictly spent in a routine.  Total time is the time spent in a
    routine and all subroutines called from that routine.}}
\end{table}
\new{The reasons for the slow-down are obvious when looking at the cost of
certain key operations, as shown in Table~\ref{tab:costs}.  Note that
more than half the time of the uniform grid calculation is the routine
\texttt{computeRHS()}.  In contrast, the amount of time in this routine
is significantly reduced for all AMR calculations, as one would expect,
since this routine will scale with the number of cells.  That the AMR
time in \texttt{computeRHS()} is no more than 24\% of the total time
suggests that the AMR cases are spending a great deal of time in AMR
overhead.  

One obvious source of overhead is regridding, the cost for which is
accounted in \texttt{regridHierarchies()}.  As expected, we see that the
AMR1 case spends the most time in regridding while the AMR3 case spends
the least.  The absolute total time spent in AMR1 regridding is more than
ninety percent of the total time the uniform mesh case spends in
\texttt{computeRHS()}.  However, for the AMR3 case, which regrids on
every eighth step, the regridding cost is much more reasonable.

The \texttt{fillData()} routine is the top level
routine that applies boundary conditions (hence the non-zero cost even
for the uniform grid case), fills patch ghost cells on the interior of
the domain, and fills new patches from patches at the same (\ie, copy)
or coarser (\ie, interpolate) mesh levels.  While \texttt{fillData()}
accounts for only 6\% of the uniform mesh calculation total time, it
represents 40-60\% of the total time for the AMR calculations.  One of
the routines that constitutes a significant portion of
\texttt{fillData()} is the finite volume WENO refinement in
\texttt{WENO\_2D()}; it can be seen that with less frequent refinement,
less time is spent in this routine.  Note that, while the absolute time
in WENO interpolation decrease monotonically from AMR1 to AMR3, the
relative times peak with AMR2; this is a trade-off between more mesh
and less frequent regridding.  Furthermore, we note that the cost of
limited, high-order interpolation for the intra-hierarchy interpolations
necessary in AMR is not a cost specific to the Vlasov-Poisson system;
an AMR method based on a higher-order discretization for any PDE system
will need to address the efficiency of such interpolations.

The other two routines shared by all four cases,
\texttt{evaluateInstantaneousConstraints()} and \texttt{reduce()}, are
provided to show that these operations spend roughly the same percentage of
time whether for the uniform mesh or for the AMR cases.  Note that
\texttt{evaluateInstantaneousConstraints()} is marginally more expensive
for all AMR cases because this routine includes the Poisson solve, which
requires more complicated reductions across the AMR hierarchies.  The
AMR1 case is more expensive in absolute time because it has more patches
to reduce. However, not that the reductions and constraint evaluations
for AMR2 and AMR3 are about the same cost as or even cheaper than (in absolute
time) the uniform case.}

\begin{table}
\centering
\new{
\begin{tabular}{|l%
 |>{\PBS\raggedleft\hspace{0pt}}p{.7in}%
  >{\PBS\raggedright\hspace{0pt}}p{.7in}%
 |>{\PBS\raggedleft\hspace{0pt}}p{.7in}%
  >{\PBS\raggedright\hspace{0pt}}p{.7in}|}\hline
\textbf{Routine} & \multicolumn{2}{|c|}{\textbf{Exclusive Time (s)}} & \multicolumn{2}{|c|}{\textbf{Total Time (s)}}\\\hline\hline
\multicolumn{5}{|l|}{\texttt{ConservativeHighOrderRefine::WENO\_2D()}}\\\hline
\textit{Mask Reduction}  & 742.6 & (27\%) & 824.3 & (30\%) \\\hline
\textit{Sub-Patch Reduction} & 642.5 & (24\%) & 727.5 & (27\%) \\\hline\hline
\multicolumn{5}{|l|}{\texttt{ReductionAlgorithm::reduce()}}\\\hline
\textit{Mask Reduction} & 27.14 & (0.99\%) & 245.7 & (9.0\%) \\\hline
\textit{Sub-Patch Reduction} & 7.505 & (0.29\%) & 157.9 & (5.9\%) \\\hline\hline
\multicolumn{5}{|l|}{\texttt{ConservativeHighOrderRefine::WENO\_1D()}}\\\hline
\textit{Mask Reduction} & 0.00 & (0\%) & 0.00 & (0\%) \\\hline
\textit{Sub-Patch Reduction} & 5.399 & (0.20\%) & 10.88 & (0.41\%) \\\hline
\end{tabular}}
\label{tab:reduction}
\caption{Comparison of the processing time of the Mask Reduction and
  Sub-Patch Reduction algorithms.  Exclusive time is the time strictly
  spent in a routine.  Total time is the time spent in a routine and
  all subroutines called from that routine.}
\end{table}
Finally, in Table~\ref{tab:reduction}, we present some timings for the
routines related to the two reduction algorithms described in
Section~\ref{sec:moment}.  \new{Note that these results were computed using
WENO interpolation and the AMR1 parameters.}  For the
Mask Reduction, we note that the \texttt{WENO\_2D} routine is called 
for intra-hierarchy regridding \new{and communication} and
  inter-hierarchy reduction calls.
With the Mask Reduction Algorithm, a great deal of time 
is spent in the 2D interpolation routine, and the reductions account for
\new{9\%} of the total run time.  For the Sub-Patch Reduction,  the
\texttt{WENO\_2D} routine is not called, so its reported time is
strictly from intra-hierarchy regridding \new{and communication} calls.
With the Sub-Patch Reduction Algorithm, \new{the time spent in the
\texttt{WENO\_2D} routine is reduced by roughly 10\%, and it is replaced
by about 1.3\% of additional work in the \texttt{WENO\_1D} routine.
The total cost of the Sub-Patch Reduction Algorithm is 64\% of the Mask
Reduction Algorithm.  The comparative performance is what was
anticipated, although, admittedly, for 1D+1V Vlasov-Poisson, the
savings are not dramatic.  Nevertheless, in higher dimensions, there
will be a more significant benefit; for 2D+2V Vlasov-Poisson, the Mask
Reduction Algorithm will require four interpolations in each cell in the
four-dimensional mesh (scaling like $N^4$), whereas the Sub-Patch
Reduction Algorithm will require only two interpolations in cell in a
two-dimensional mesh (scaling like $N^2$).}

\section{Conclusions}
\label{sec:conclusions}
We have demonstrated the application of block structured adaptive mesh
refinement to the 1D+1V Vlasov-Poisson system as implemented in the
\Valhalla code based on the SAMRAI AMR library.  The primary
complication comes from a solution state comprised of variables of
different dimensions. The considerations and algorithms required to
extend standard single-dimensional block structured AMR have been
presented. In particular, algorithms for reduction and injection
operations that transfer data between mesh hierarchies of different
dimensions were explained in detail.  In addition, modifications to the
basic AMR algorithm due to our use of high-order spatial and temporal
discretizations were presented.  Preliminary results for a standard
Vlasov-Poisson test problem were presented, and these results indicate
that there is potential for savings\new{, both in memory and in compute
time,} for at least some Vlasov problems.  The effectiveness for any
particular problem will depend intimately on the features of the solution.

There are several obvious directions for future work.  Currently, we are
working on generalizing the \Valhalla code to 2D+2V and higher
dimensions.  The SAMRAI library is quite general and supports arbitrary
dimension.  Moving to   
4D calculations and beyond opens up several new directions for
investigation.  When the configuration space problem is in 2D or 3D, there
is potential for savings from adaptivity in configuration space.  It is
straightforward to enable this generalization, but it is unclear if the
additional cost of the necessary FAC algorithm and of the AMR overhead
will justify the complication, particularly when the solution time will
be dominated by operations in the 4D or higher phase space.  With larger
phase-space problems, an efficient parallel decomposition will be
necessary; we have already indicated potential advantages of providing
each species a distinct subset of processors, but empirical results are
needed.  It also will prove beneficial to allow for asynchronous time
stepping in the AMR advancement; an example of the necessary
modifications to the time advancement algorithm has been shown~\cite{McCo10}.

Dimensions above three also require additional empirical investigation
for AMR efficiency.  As indicated earlier, the potential savings from
AMR increases geometrically with dimension, but AMR overhead, much of
which scales with the number of cells at coarse-fine boundaries, also
increases with dimension.  Whether the overhead costs in 4D and above
negate the savings remains an open issue and must be the subject of
future studies.

AMR overhead, even in lower dimensions, still requires further reduction.
One clear path is to make use of hybrid parallelism as multicore
architectures become more prevalent.  Much of the
computations contributing to AMR overhead are sequentially executed on
a node but are completely independent and thus ideal for task parallelism.
\new{I}mplementations should also be found that further optimize the
conservative, high-order intra-hierarchy interpolations.

Finally, the topic of refinement criteria requires further
investigation.  Heuristic or \emph{a posteriori} error indicators that
seek to minimize local error are sufficient but not optimal, depending
on the desired results of the calculation.  Most quantities of
interest exist in configuration space, that is, the macroscopic
quantities like temperature and density that can be readily measured in
the laboratory.  The reductions leading to configuration space
quantities integrate out  much of the finer details in phase space,
which suggests that it may be inefficient to resolve all of the finer
phase-space structure.  Future investigations should consider (i) the phase
space resolution requirements to obtain accurate configuration space
quantities of interest and (ii) whether more efficient phase-space refinement
criteria can be formulated based on these configuration-space accuracy
requirements.

\section*{Acknowledgements}
The authors would like to thank Dr. Bruce Cohen, Dr. Richard Berger, and
Dr. Stephan Brunner for their many helpful comments and suggestions.

This work was performed under the auspices of the
U.S. Department of Energy by Lawrence Livermore National Laboratory
under contract number DE-AC52-07NA27344.  This work was funded by the
Laboratory Directed Research and Development Program at LLNL under
project tracking code 08-ERD-031. LLNL-JRNL-515291-DRAFT.

\bibliography{journalISI20,publishers,vlasov}
\bibliographystyle{unsrtnat}

\end{document}